\documentclass[onecolumn,preprint,preprintnumbers,nofootinbib,floats,12ptm,tightenlines]{revtex4}
\pdfoutput=1

\usepackage{amsmath,amssymb,color,mathrsfs,verbatim,epsfig, wasysym}
\usepackage[hyperfootnotes=false]{hyperref}
\usepackage{slashed}
\usepackage{multirow}
\usepackage{xspace}
\usepackage{ulem}
\allowdisplaybreaks

\begin{document}

\baselineskip=18pt

%%%%%%%%%%
%%%%%%%%%%    Title page
%%%%%%%%%%

\thispagestyle{empty}
\vspace{20pt}
\font\cmss=cmss10 \font\cmsss=cmss10 at 7pt

\begin{flushright}
\small 
\end{flushright}

\hfill
\vspace{20pt}

\begin{center}
{\Large \textbf
{
Cosmological Relaxation\\[10pt] from Dark Fermion Production
}}
\end{center}

\vspace{15pt}
\begin{center}
{ Kenji~Kadota$^{\, a}$, Ui~Min$^{\, b}$, Minho~Son$^{\, b}$ and Fang~Ye$^{\, b}$}
\vspace{30pt}

$^{a}$ {\small \it Center for Theoretical Physics of the Universe, Institute of Basic Science (IBS), \\ Daejeon 34126, Republic of Korea 
}
\vskip 3pt
$^{b}$ {\small \it Department of Physics, Korea Advanced Institute of Science and Technology, \\ 291 Daehak-ro, Yuseong-gu, Daejeon 34141, Republic of Korea
}

\end{center}

\vspace{20pt}
\begin{center}
\textbf{Abstract}
\end{center}
\vspace{5pt} {\small

We consider the cosmological relaxation solution to the electroweak hierarchy problem using the fermion production as a dominant friction force. In our approach, neither super-Planckian field excursions nor a large number of e-folds arise, and scanning over thermal Higgs mass squared is avoided. The produced fermions from the relaxion source through the derivative coupling are SM-singlets, what we call dark fermions, and they can serve as the keV scale warm dark matter candidates.}

\vfill\eject
\noindent

\tableofcontents
\newpage

%%%%%%%%%%%%%%%%%%%%%%%%%%%%%%%%%%%%%%%%
%%%%%%%%%%%%%%%%%%%%%%%%%%%%%%%%%%%%%%%%
%%%%%%%%%%%%%%%%%%%%%%%%%%%%%%%%%%%%%%%%
%%%%%%%%%%%%%%%%%%%%%%%%%%%%%%%%%%%%%%%%
%%%%%%%%%%%%%%%%%%%%%%%%%%%%%%%%%%%%%%%%
%%%%%%%%%%%%%%%%%%%%%%%%%%%%%%%%%%%%%%%%
%%%%%%%%%%%%%%%%%%%%%%%%%%%%%%%%%%%%%%%%
\section{Introduction}
\label{sec:intro}

Despite the Higgs discovery at the LHC~\cite{Aad:2012tfa,Chatrchyan:2012xdj} and the continuous measurements of its properties, the smallness of the Higgs mass still remains mysterious within the quantum field theory of the Standard Model (SM). The traditional approach to the naturalness problem such as the supersymmetry~\cite{Fayet:1976aa,Fayet:1977,Fayet:1979aa,Farrar:1978aa} or Higgs compositeness~\cite{Kaplan:1983fs,Kaplan:1983sm,Georgi:1984ef,Georgi:1984af,Dugan:1984hq,Contino:2003ve} relied on the symmetry-based selection rules and the light new particles such as top partners in minimal scenarios~\cite{Dimopoulos:1995mi,Cohen:1996vb} were expected to be observed at the LHC. However, the absence of the evidence for the New Physics near the electroweak scale so far only degrades the naturalness principle as a valid guiding principle that has been used for decades to extend the SM~\cite{Dine:2015xga}. 

The relaxation meachanism is a recent new approach to the naturalness problem based on the cosmological dynamics of the Higgs mass squared~\cite{Graham:2015cka}. In the relaxation approach, the smallness of the Higgs mass parameter is not associated with the symmetry, and in principle no new physics near the electroweak scale is required to show up.  After the first realization of the relaxation mechanism for the electroweak hierarchy problem in~\cite{Graham:2015cka}, a series of improvements attempting to resolve downsides in the original scenario have appeared~\cite{Espinosa:2015eda,Hardy:2015laa,Patil:2015oxa,Antipin:2015jia,Jaeckel:2015txa,Gupta:2015uea,Batell:2015fma,Matsedonskyi:2015xta,Marzola:2015dia,DiChiara:2015euo,Ibanez:2015fcv,Hook:2016mqo,Fonseca:2016eoo,Fowlie:2016jlx,Evans:2016htp,Kobayashi:2016bue,Choi:2016luu,Flacke:2016szy,McAllister:2016vzi,Lalak:2016mbv,Choi:2016kke,Evans:2017bjs,Beauchesne:2017ukw,Batell:2017kho,Nelson:2017cfv,Davidi:2017gir,Fonseca:2017crh,Tangarife:2017rgl,Ibe:2019udh,Gupta:2019ueh}. Among them, the one using the particle production~\cite{Hook:2016mqo} is of particular interest. We postpone the overview of the original idea and its variant using the particle production to Section~\ref{sec:RO}.

In this work, we newly investigate the cosmological relaxation solution to the electroweak hierarchy problem using the particle production of fermions.
The application of the fermion production~\cite{Greene:1998nh,Greene:2000ew} in the Beyond the SM (BSM) scenarios has been somewhat limited mainly due to the Pauli-blocking (unlike the parametric resonance for scalars~\cite{Landau:1991wop} or the particle production of the tachyonic gauge bosons~\cite{Anber:2009ua}). A recent interesting application in cosmology is the axion inflation in~\cite{Adshead:2015kza,Adshead:2018oaa} where the backreaction from the fermion production was shown to be more efficient than the dissipation via the Hubble friction in supporting the slow-roll of the inflaton and from where we adopted many technical results. The goal of this work is to investigate the plausibility of realizing the cosmological relaxation using the fermion production as a dominant way of dissipating relaxion energy while aiming to maintain the cutoff scale in a similar size to that of other variant models. Two benchmark BSM scenarios that we consider in this work are the non-QCD model in~\cite{Graham:2015cka} and the double scanner mechanism proposed in~\cite{Espinosa:2015eda}. While the non-QCD model does not look satisfactory (not conclusive though), we use it as a toy example for the simpler demonstration of the underlying physics, and the double scanner mechanism will be taken as our proof-of-concept example.  We will show that our proof-of-concept example avoids downsides in the original relaxation scenario, but at the same time it brings new types of theoretical challenges. We will also demonstrate that the fermions have to be SM singlet for the mechanism to work and their masses are interestingly in the right ballpark to be the dark matter candidates.

The paper is organized as follows. In Section~\ref{sec:RO}, we review the original relaxation mechanism in~\cite{Graham:2015cka} and some of its improvements. In Section~\ref{sec:model}, we survey two BSM models, namely non-QCD model and double scanner model, focusing on whether the backreaction from the fermion production can support a slow-rolling relaxion while satisfying all theoretical constraints. In Section~\ref{sec:DM}, we discuss about the prospect for dark matter candidates and check their compatibility with the current phenomenological and astrophysical bounds. The concluding remarks are summarized in Section~\ref{sec:con}. In Appendix~\ref{app:FP}, we provide all detailed derivations of our analytic expressions that we have used throughout our work.

%%%%%%%%%%%%%%%%%%%%%%%%
%%%%%%%%%%%%%%%%%%%%%%%%
%%%%%%%%%%%%%%%%%%%%%%%%
\section{Ralaxation overview}
\label{sec:RO}

In this section, we briefly review the cosmological relaxation of the electroweak scale proposed by Graham, Kaplan and Rajendran~\cite{Graham:2015cka} (which we refer to as GKR) as well as the issues in the GKR scenario. Relaxion is an axion-like particle (ALP) whose discrete shift symmetry is softly and explicitly broken by a small coupling. The relaxion potential in GRK scenario takes the form
\begin{equation}\label{eq:GKR:pot}
\begin{split}
 \Delta V =&\  \left ( -\Lambda^2 + g \phi \right ) |h|^2 + \left ( g \Lambda^2 \phi + \cdots \right ) + \Lambda_c^4 \cos \left (\phi/f \right ),
\end{split}
\end{equation}
where $\phi$ is the relaxion field, $h$ is the Higgs doublet, $\Lambda$ is the cutoff scale of the relaxion model, and $g$ is a small dimensionful coupling. The ellipsis in Eq.~(\ref{eq:GKR:pot}) refers to the higher order terms in $g\phi/\Lambda^2$. The small potential terms for $\phi$ are technically natural in the sense that the discrete shift symmetry $\phi \to \phi + 2\pi f$ is restored when the coupling $g\to 0$. 
The periodic potential $\Lambda_c^4 \cos \left (\phi/f \right )$ arises from model-dependent non-perturbative dynamics (either SM QCD or non-QCD strong gauge group), and the height of the potential barriers $\Lambda_c^4\propto M^{4-n} v^n$, where the integer $n\in [1,\,4]$~\footnote{
An example with $n=0$ can be found in~\cite{Hook:2016mqo}, whereas a scenario with the QCD axion (non-QCD axion) is an example with $n=1$ ($n=2$)~\cite{Graham:2015cka}.}, $M$ is a parameter of mass dimension, and $v$ is the Higgs vacuum expectation value (VEV). The QCD relaxion is problematic since it generates an $\mathcal O(1)$ shift in the $\theta$-term, causing the strong CP problem. Either an additional mechanism (see~\cite{Graham:2015cka,Nelson:2017cfv} for example) must be introduced to remedy the problem or one needs to consider a new strong gauge group instead of QCD.

The relaxion field initially can be anywhere, while $\phi \gtrsim \Lambda^2/g$, such that the effective Higgs mass squared is positive, $\mu^2\equiv -\Lambda^2 + g \phi>0$, and electroweak symmetry is unbroken. Since the cosine potential is switched off in the unbroken phase, the evolution of the relaxion is driven by the linear potential. The mechanism is insensitive to the initial value of $\phi$ since the relaxion is slow-rolling due to the Hubble friction. The relaxion must scan $\mathcal{O}(1)$ fraction of the field space to naturally pass a critical point $\phi_c = \Lambda^2/g$ where $\mu^2$ changes its sign, and the Higgs develops a nonvanishing VEV.  The increasing Higgs VEV backreacts onto the potential by growing the potential barriers. Eventually it results in a compensation between the slope of the periodic potential $\Lambda_c^4/f$ and the linear slope $g\Lambda^2$, stabilizing the relaxion in a local minimum near a small value, namely, the electroweak scale $v$. Therefore, the smallness of the Higgs VEV is explained by the dynamical evolution of the relaxion field instead of fine-tuning or anthropics. 

Certain conditions must be satisfied to make the relaxation mechanism work. The slow-roll of the relaxion must be long enough such that it can scan an $\mathcal{O}(1)$ fraction of the whole field range, and it sets a lower bound on the number of e-folds $N_e \gtrsim H^2/g^2$: $\Delta \phi\sim \dot\phi \Delta t \sim \dot\phi\, N_e/H \sim ( g\Lambda^2/H^2)\, N_e \gtrsim \Lambda^2/g$ where we have used the slow-roll condition $3H\dot \phi+\frac{d\Delta V}{d\phi}\sim 0$. The vacuum energy should be greater than the typical relaxion energy density, $H^2M_P^2\gtrsim \Lambda^4$. The potential barrier must be formed within the Hubble sphere, $H^{-1}>\Lambda^{-1}$. In addition, it has to be the classical rolling instead of the quantum spreading that sets the vacuum into the correct one, leading to $H^2/{\dot\phi}<1$. After the relaxion stops rolling and the reheating of the universe occurs, a sufficiently high temperature may erase the barrier and cause the relaxion to roll again. Hence, either the reheating temperature must be low enough such that the barrier does not get melt or the traveling distance during the second rolling must be short enough not to overshoot the electroweak scale.

Some downsides, however, exist for the original GKR scenario which relies on the Hubble expansion to dissipate the relaxion energy. The typical size of the dimensionful coupling $g$ is tiny (e.g. $g\sim 10^{-31}$ GeV for the QCD axion model), and it leads to the super-Planckian relaxion scanning during the inflation and the exponentially long e-folding. The fact that those downsides are linked to the inflation suggests looking for a more efficient way of dissipating the relaxion energy than via the Hubble friction. Such an example is found in the particle production sourced by the relaxion field. In Ref.~\cite{Hook:2016mqo}, Hook and Marques-Tavares (HMT) utilized the particle production of tachyonic gauge bosons through $\phi F\tilde{F}$ as an efficient way of dissipating relaxion energy. The exponential production of the electroweak gauge boson naturally occurs to stop the relaxion at a value that generates the electroweak scale. Since the mechanism can work in a non-inflationary era, the issues mentioned above can be avoided. However, unlike the GKR scenario, the original HMT mechanism operates in the broken phase of the electroweak symmetry.
Such a model requires a specific, usually non-trivial  UV completion, e.g. from a left-right symmetric model as in the appendix of~\cite{Hook:2016mqo}. The cutoff scale of the HMT scenario is relatively low, up to $10^{4\sim 5}$ GeV, and hence this scenario addresses the little hierarchy problem (see also~\cite{Fonseca:2018xzp,Son:2018avk} for more recent and updated analyses on the HMT mechanism).

%%%%%%%%%%%%%%%%%%%%%%%%
%%%%%%%%%%%%%%%%%%%%%%%%
%%%%%%%%%%%%%%%%%%%%%%%%
\section{Ralaxation from dark fermion production}
\label{sec:model}
\begin{figure}[tph]
\begin{center}
\includegraphics[width=0.60\textwidth]{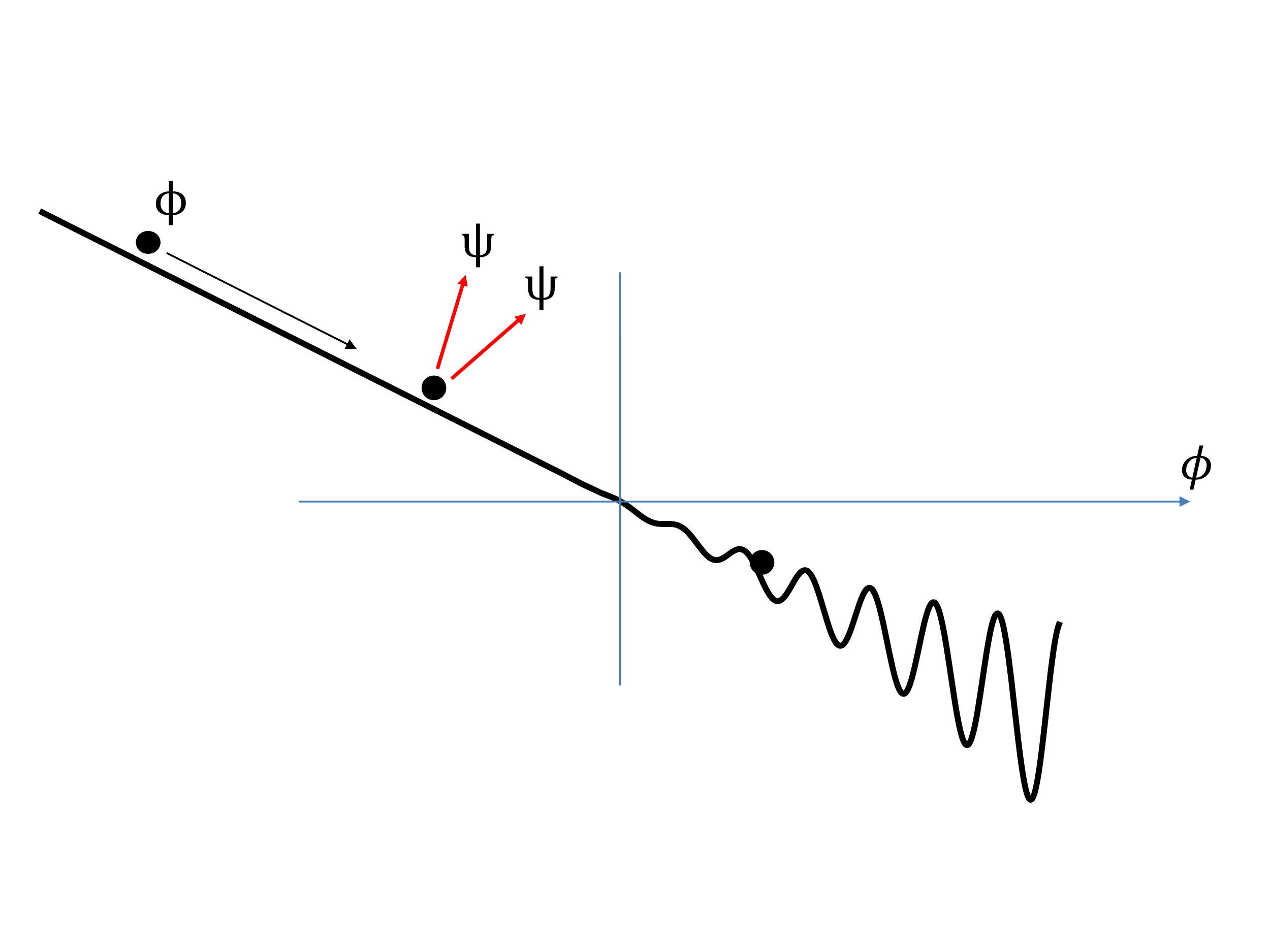}\quad
\caption{\small A cartoon of the relaxation with the dark fermion production.}
\label{fig:fermionProduction}
\end{center}
\end{figure}

In this section, we investigate the plausibility of the fermion production as an alternative to the Hubble friction as a dominant way of dissipating relaxion energy to achieve the slow-rolling of the relaxion. The cartoon picture for the situation is illustrated in Fig.~\ref{fig:fermionProduction}. As in the original GKR scenario, we rely on the cosine potential being switched on only in the broken phase of the electroweak symmetry to stop the relaxion at the right place while a large field excursion of the relaxion occurs in the unbroken phase. Since the fermion production can not be exponential due to Pauli-blocking (unlike the case of the tachyonic gauge boson), it would be difficult to be implemented to work in a similar manner to~\cite{Hook:2016mqo} in the broken phase as a way of naturally selecting the electroweak scale. 

%%%%%%%%%%%%%%%%%%%%%%
%%%%%%%%%%%%%%%%%%%%%%

%%%%%%%%%%%%%%%%%%%%%%
%%%%%%%%%%%%%%%%%%%%%%
\subsection{Dark fermion production}
The fermionic system that couples to the relaxion field $\phi$ through the derivative coupling in an expanding Universe can be written as
\begin{equation}\label{eq:action:rot}
 \Delta\mathcal {S} = \int d^4 x \sqrt{-g} \Big [ \bar{\psi}\left ( i e^\mu_{\ a} \gamma^a D_\mu - m_\psi - \frac{1}{f_\psi} e^\mu_{\ a} \gamma^a\gamma^5\partial_\mu \phi \right ) \psi  \Big ]~,
\end{equation}
where $e^\mu_{\ a}$ is a vierbein and $D_\mu$ is the covariant derivative due to the spin connection from the scale factor $a$ in the metric,
\begin{equation}
  ds^2 = dt^2 - a^2 d {\bf x}^2 =  a^2 \left ( d\tau^2 - d {\bf x}^2 \right )~.
\end{equation}
The overall dependence on the scale factor $a$ in the fermionic system can be removed via rescaling, $\psi \rightarrow a^{-3/2}\psi$, after which the Lagrangian density for the fermions in the conformal time takes a simple form,
\begin{equation}\label{eq:deri:coupling}
 \Delta\mathcal {L} = \bar\psi \left ( i \gamma^\mu \partial_\mu - m_\psi a  - \frac{1}{f_\psi} \gamma^\mu \gamma^5 \partial_\mu \phi\right ) \psi ~.
\end{equation}
Throughout this work, we will assume that the relaxion field $\phi$ is spatially homogeneous.  As was pointed out in~\cite{Adshead:2018oaa} (see~\cite{Min:2018rxw} for the related discussion), the fermion production in the basis with the derivative coupling as in Eq.~(\ref{eq:deri:coupling}) is not accurately estimated. A prescription is to go to a new basis, where the fermion production is unambiguously estimated, via the rotation~\cite{Adshead:2018oaa}
\begin{equation}\label{eq:rotToinertial}
  \psi \rightarrow e^{-i\, \gamma^5 \phi/f_\psi} \psi~.
\end{equation}
In the new basis, the interaction between the relaxion and fermions takes the form,
\begin{equation}\label{eq:nonderi:coupling}
  \Delta{\mathcal{L}} = \bar{\psi} \left (i \gamma^\mu \partial_\mu - m_R + i\, m_I \gamma^5 \right ) \psi~,
\end{equation}
where $m_R = m_\psi\, a\, \cos(2\phi/f_\psi)$ and $m_I = m_\psi\, a\, \sin(2\phi/f_\psi)$. One clearly sees in Eq.~(\ref{eq:nonderi:coupling}) that the fermion production vanishes in the massless limit since the fermions become free fields. The new basis is more suitable to define the fermion number mainly because the Hamiltonian for the fermion takes a simple quadratic form,
\begin{equation}
\mathcal{H} = \bar{\psi} \left ( - i \gamma^i \partial_i + m_R - i\, m_I \gamma^5 \right ) \psi~,
\end{equation}
where $\psi$ is taken to be a quantum field in terms of the creation and annihilation operators.
The fermion number density for a momentum $\vec{k}$ and helicity $r$ is given by the vacuum expectation value of the number operator at a finite time, namely $\langle 0 | a^\dagger_r (\vec{k}) a_r (\vec{k}) | 0 \rangle$. The detailed computation will be given in Appendix~\ref{app:FP}. Here, we simply take the final result to discuss about the main feature of the scenario. The fermion $\psi$ in Eq.~(\ref{eq:deri:coupling}) can not be a SM one since the SM fermion will be massless during the scanning era in the unbroken phase of the electroweak symmetry. We assume that $\psi$ is a SM singlet massive Dirac fermion (what we call the dark fermion in the following) and that its mass is a Higgs VEV independent~\footnote{Another justification for non-SM fermions is that the SM fermions, once produced, might quickly thermalize, preventing from scanning the zero-temperature Higgs mass squared, and the issue can be avoided for dark fermions.}. The non-SM fermion $\psi$ could be a candidate for a dark matter as a bonus. 
We will investigate its plausibility in detail in Section~\ref{sec:DM}.

%%%%%%%%%%%%%%%%%%%%%%
%%%%%%%%%%%%%%%%%%%%%%
\subsection{Slow-rolling from backreaction}
In presence of the backreaction due to the fermion production, the equation of motion for the relaxion reads
\begin{equation}\label{eq:eom:phi}
  \ddot{\phi} + 3 H \dot{\phi} + V_\phi(\phi) = \mathcal{B}~,
\end{equation}
where $V_\phi \equiv \partial V/\partial\phi$ and dot denotes the differentiation with respect to the cosmic time, for instance, $\dot\phi = \partial_t \phi$. The backreaction $\mathcal{B}$ from the fermion production in Eq.~(\ref{eq:eom:phi}) is given by
\begin{equation}
\mathcal{B} = \frac{2m_\psi}{a^3 f_\psi} \langle \bar{\psi} \big [ \sin \left (2\phi/f_\psi \right ) + i\, \gamma^5 \cos  \left (2\phi/f_\psi \right )  \big ] \psi \rangle~.
\end{equation}
The exact evaluation of the backreaction $\mathcal{B}$ can be found in Appendix~\ref{app:FP}.  In the limit $\mu^2 \equiv m^2_\psi/H^2 \ll \xi$ and $\xi \gg 1$, it is approximated to be
\begin{equation}
 \mathcal{B} \sim  - \frac{1}{f_\psi} H^4 \mu^2 \xi^2~,
\end{equation}
where $\xi$ is defined as the ratio of the velocity of $\phi$ to the scale $f_\psi$ in the Hubble unit,
\begin{equation}
  \xi \equiv \frac{1}{2H} \frac{\dot{\phi}}{f_\psi}~.
\end{equation}
The parameter $\xi$ plays a crucial role in controling the slow-rolling of the relaxion, or the strength of the fermion production or energy dissipation. A strong fermion production occurs when the adiabatic condition is strongly violated. This implies that the relaxion should maintain a sizable velocity to strongly depart from the adiabaticity. What controls the size of the velocity of $\phi$ would be the slope of the linear potential, namely $g$ in our framework. Having said that, the fermion production assisted slow-rolling favors a bigger slope than that in the original GKR relaxation operating with the Hubble friction, whereas the slope is smaller compared to the case of the inflation through the fermion production in a steep axionic potential~\cite{Adshead:2018oaa}. 

While the velocity of the slow rolling $\phi$ during the inflation in the absence of the backreaction is determined by equating the second and third terms on the left hand side of Eq.~(\ref{eq:eom:phi}), the velocity in our scenario is engineered to be determined by equating $V_\phi(\phi)$ with the backreaction term $\mathcal{B}$ in Eq.~(\ref{eq:eom:phi}):
\begin{equation}\label{eq:slowroll:phidot}
  V_\phi(\phi) (= g \Lambda^2) \sim \mathcal{B} \quad \rightarrow \quad \dot{\phi} \sim 2\, \frac{g^{1/2} \Lambda f_\psi^{3/2} }{m_\psi} \sim {\rm constant}~.
\end{equation}
It implies that the size of the backreaction is solely determined by the size of the slope $g$ as the left hand side of the slow roll equation in Eq.~(\ref{eq:slowroll:phidot}) depends only on the slope for a given cutoff scale $\Lambda$.

%%%%%%%%%%%%%%%%%%
\subsection{Theoretical constraints for relaxation}
\label{sec:theory:constraint}

In the following, we list various constraints for the successful relaxation from the dark fermion production, and we express them as a lower or upper bound on the dark fermion mass. 
\begin{enumerate}
\item Consistency of slow rolling:\\
The friction term due to the Hubble parameter should be subdominant in order for the slow rolling to be maintained by the backreaction, or to be consistent with the slow-rolling condition in Eq.~(\ref{eq:slowroll:phidot}),
\begin{equation}\label{eq:constraint:consist}
  V_\phi (\phi) > 3 H \dot{\phi} \quad \rightarrow \quad m_\psi > 6\, \frac{H}{\Lambda}\frac{f_\psi^{3/2}}{g^{1/2}}~.
\end{equation}
\item Validity of EFT:\\
The validity of the EFT demands
\begin{equation}\label{eq:constraints:EFT}
  \dot{\phi} \lesssim \Lambda^2 \quad \rightarrow \quad m_\psi \gtrsim 2\, \frac{g^{1/2} f^{3/2}_\psi}{\Lambda}~.
\end{equation}
\item Small dark fermion energy density:\\
The energy density carried by dark fermions must be smaller than the total energy,
\begin{equation}\label{eq:constraint:fermionE}
  \rho_\psi \sim 16 \pi^2 H^4 \mu^2 \xi^3 \lesssim H^2 M_p^2 \quad \rightarrow \quad m_\psi \gtrsim \frac{\Lambda^3}{H^3} \frac{g^{3/2} f_\psi^{3/2}}{M^2_p}~,
\end{equation}
where the expression for the fermion energy density holds only in the approximation $\mu^2 \ll \xi$ and $\xi \gg 1$.
\item Small kinetic energy:\\
The kinetic energy needs to be smaller than the total energy
\begin{equation}\label{eq:constraint:kin}
  \dot{\phi}^2 \lesssim H^2 M_p^2 \quad \rightarrow \quad m_\psi \gtrsim 2\, \frac{\Lambda}{H}\frac{g^{1/2}f_\psi^{3/2}}{M_p}~,
\end{equation}
which is automatically satisfied if the condition in Eq.~(\ref{eq:constraints:EFT}) is satisfied, provided that the typical energy density of the relaxion should be smaller than the total energy density, $\Lambda^4 \lesssim H^2 M^2_p$.
\item (Sub-Planckian) Large field excursion:\\
While the natural scanning process requires a large field excursion $\Delta\phi \gtrsim \Lambda^2/g$, and it can set a lower bound on the number of e-folding through $\Delta\phi = \dot{\phi} \Delta t = \dot{\phi}\, (N_e/H) \gtrsim \Lambda^2/g$, we express the constraint as an upper bound on $m_\psi$ for a fixed $N_e$,
\begin{equation}\label{eq:constraint:largephi}
  \Delta\phi \gtrsim \frac{\Lambda^2}{g}    \quad \rightarrow \quad m_\psi \lesssim 2N_e\, \frac{g^{3/2} f_\psi^{3/2}}{H \Lambda}~.
\end{equation}
In this work, we will consider a modest size of the e-folding, $N_e \sim \mathcal{O}(10^{1\sim 3})$. It is consistent with the viable parameter space according to our numerical simulation. It also guarantees the validity of our analysis based on the analytic formula obtained using approximated solutions of the equations of motion as it requires $N_e \gtrsim \mathcal{O} (10)$ (see Appendix~\ref{app:sec:eom} for the detail). On the other hand, requiring the field excursion to be sub-Planckian leads to a lower bound on $m_\psi$, 
\begin{equation}
  M_p > \Delta \phi \quad \rightarrow \quad m_\psi > 2 N_e \frac{\Lambda}{H} \frac{g^{1/2} f_\psi^{3/2}}{M_p}~,
\end{equation}
and it becomes weaker than the constraint in Eq.~(\ref{eq:constraints:EFT}) when $\Lambda^2 < H M_p/N_e$ is satisfied.
\item Classical rolling beats quantum spreading:\\
The evolution of $\phi$ should be dominated by the classical rolling over the quantum spreading,
\begin{equation}\label{eq:classrolling:constraint}
  \dot{\phi} \Delta t \gtrsim H \quad \rightarrow \quad m_\psi \lesssim \frac{g^{1/2} f_\psi^{3/2} \Lambda}{H^2} ~.
\end{equation}
\item Barriers form:\\
The barrier should form within the Hubble scale, $H \lesssim \Lambda_c$.
\item Precision of mass scanning:\\
The effective Higgs mass squared should be selected with the enough precision not to overshoot the electroweak scale, and it requires
\begin{equation}
  \Delta m^2_h \sim g\Delta \phi \sim g\, 2 \pi f \lesssim m^2_h~,
\end{equation}
which is easily satisfied in the parameter space of interest.
\end{enumerate}
In addition to the constraints listed above, when the relaxion is
scanning over the effective Higgs mass squared, the temperature is required to be negligible compared to the electroweak scale, not to scan over the thermal Higgs mass squared. We help this issue by considering our relaxation mechanism during inflation driven by a separate inflaton sector (where the inflationary Hubble $H$ is much higher than $\Lambda^2/M_P$). The constraint can be avoided when the interaction rate between the relaxion and the SM sector can be made small enough to be diluted during the inflation~\footnote{We find that this is plausible only in the double scanner mechanism among two scenarios in Sections~\ref{sec:nonQCD} and~\ref{sec:doublescanner}. The dark fermions may thermally produce relaxions even during the inflation (it is a characteristic feature of the derivative coupling that induces strong backreaction at the higher energy or temperature) in both scenarios. However, the small non-derivative couplings between the relaxions and the SM sector in the double scanner mechanism lead to an inefficient interaction rate than $H$  during the inflation. A similar suppression is not obvious in non-QCD model as the relaxion may thermally produce new non-QCD gauge bosons $G'$ above the confinement scale through the derivative coupling $\phi G'\tilde G'$, and then the new gauge boson will thermalize the SM sector.}.

The cosine potential for $\phi$ is turned on when $\phi$ passes the critical point $\phi_c$ from which it enters into the broken phase of the electroweak symmetry. The relaxion is being trapped in one of minimum when the slope of the linear potential is balanced with the slope of the cosine potential,
\begin{equation}\label{eq:stopping:phi}
   g\Lambda^2  = \frac{\Lambda^4_c}{f}~.
\end{equation}

\begin{figure}[tph]
\begin{center}
\includegraphics[width=0.46\textwidth]{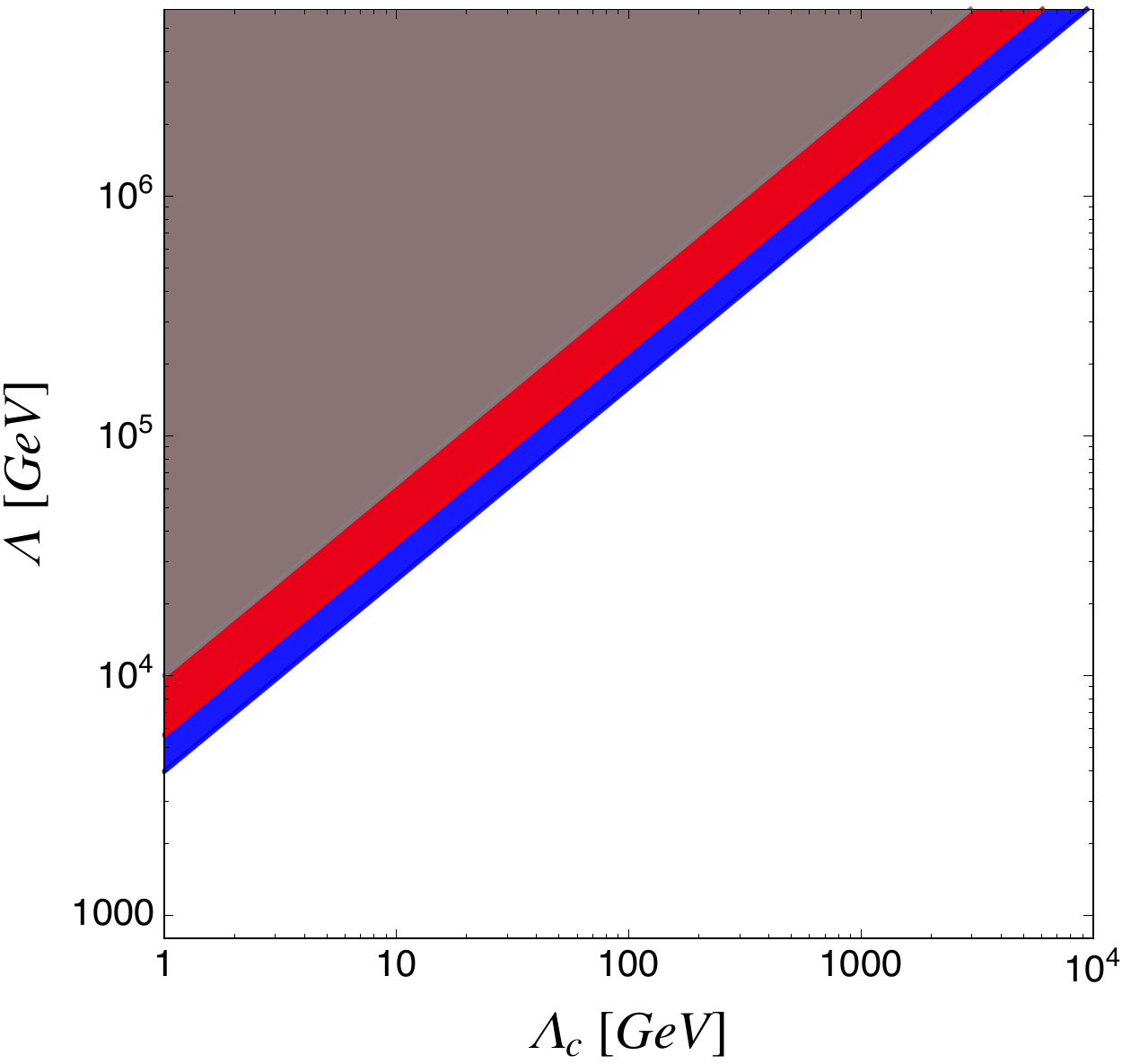}
\caption{\small The excluded region for the cutoff scale $\Lambda$ as a function of $\Lambda_c$ for $N_e = 10^2$. The red (gray) excluded region corresponds to the first (third) term in Eq.~(\ref{eq:ubound:cutoff}) for $N_e = 10^2$. The discrepancy between two excluded regions is reduced with decreasing value of $N_e$. The blue region corresponds to the case in Eq.~(\ref{eq:constraint:subMp}).}
\label{fig:excl:cutoffVSlamc}
\end{center}
\end{figure}

By combining two lower bounds on $m_\psi$ in Eqs.~(\ref{eq:constraint:consist}) and~(\ref{eq:constraints:EFT}) and two upper bounds on $m_\psi$ in Eqs.~(\ref{eq:constraint:largephi}) and~(\ref{eq:classrolling:constraint}), we can obtain four inequalities where $f_\psi$-dependence completely drops out. The parameter $g$ can be traded for $\Lambda$, $\Lambda_c$, and $f$ via the relation in Eq.~(\ref{eq:stopping:phi}). Assuming a modest size of $N_e$ and using two inequalities, $f > \Lambda$ and $H > \Lambda^2/M_p$, we can get meaningful exclusion for the cutoff as a function of $\Lambda_c$:
\begin{equation}\label{eq:ubound:cutoff}
 \Lambda < {\rm min} \big [ \left ( N_e/3 \right )^{1/10} M_p^{1/5} \Lambda_c^{4/5}~,\ \left ( 1/6 \right )^{1/7} M_p^{3/7} \Lambda_c^{4/7}~,\ N_e^{1/5} M_p^{1/5} \Lambda_c^{4/5}  \big ]~,
\end{equation}
where three bounds from the left are obtained by combining Eqs.~(\ref{eq:constraint:consist}) and~(\ref{eq:constraint:largephi}), 
Eqs.~(\ref{eq:constraint:consist}) and~(\ref{eq:classrolling:constraint}), and Eqs.~(\ref{eq:constraints:EFT}) and~(\ref{eq:constraint:largephi}), respectively. The remaining combination gives a trivial constraint. Eq.~(\ref{eq:constraint:fermionE}) does not lead to the constraint that fits to the form in Eq.~(\ref{eq:ubound:cutoff}). Using the inequality $f > \Lambda$ in Eq.~(\ref{eq:stopping:phi}) leads to $g<\Lambda^4_c/\Lambda^3$, and it can be combined with the condition for sub-Planckian field excursion $\Delta \phi \sim \Lambda^2/g < M_p$ to obtain another upper bound on the cutoff,
\begin{equation}\label{eq:constraint:subMp}
   \Lambda < M_p^{1/5} \Lambda_c^{4/5}~.
\end{equation}
The excluded region in ($\Lambda_c,\, \Lambda$) plane from Eq.~(\ref{eq:ubound:cutoff}) for $N_e = 10^2$ is illustrated in Fig.~\ref{fig:excl:cutoffVSlamc} where the strongest bound corresponds to the first one (red region) in Eq.~(\ref{eq:ubound:cutoff}) followed by the third (blue region) as next strongest, and the second one is the weakest (not shown). We also show in Fig.~\ref{fig:excl:cutoffVSlamc} the case from Eq.~(\ref{eq:constraint:subMp}) (gray region) which roughly matches to Eq.~(\ref{eq:ubound:cutoff}) for $N_e = 1$.  
%%%%%%%%%%%%%%%%%%%%%%
%%%%%%%%%%%%%%%%%%%%%%
As is evident in Fig.~\ref{fig:excl:cutoffVSlamc}, the higher barrier is more compatible with the higher cutoff scale.

In the next section, we will survey a few new physics models that predict different forms of $\Lambda_c$, and we will determine the viable parameter space consistent with all the constraints in those models. While one could explore the parameter space with the maximally allowed cutoff scale within the ballpark for $\Lambda_c$ in a specific model, throughout our work, we will fix the cutoff scale to $\Lambda \sim 10^{4\sim5}$ (aiming to address only the little hierarchy problem) as our benchmark point. Note that another type of relaxation scenario with the particle production in~\cite{Hook:2016mqo} also has a similar low cutoff scale.

%%%%%%%%%%%%%%%%%%%%%%%%
%%%%%%%%%%%%%%%%%%%%%%%%
%%%%%%%%%%%%%%%%%%%%%%%%
\subsection{Non-QCD model}
\label{sec:nonQCD}

The model that we try first as an illustration is the non-QCD model~\cite{Graham:2015cka} supplemented by dark fermions. New massive fermions $L,\, N$ (and their conjugates $L^c$, $N^c$) in the non-QCD model are charged under the new gauge group that gets strongly coupled in the low energy scales, and they have Yukawa-type couplings with the Higgs field:
\begin{equation}\label{eq:nonQCD:mass:Yukawa}
  \Delta \mathcal{L}_{non{\text -}QCD} = m_L LL^c + m_N NN^c + yhLN^c + \tilde{y}h^\dagger L^c N~,
\end{equation}
where $L$ and $N$ have the same electroweak quantum numbers as those of the lepton doublet and right handed neutrino.
The $L$ and $L^c$ fermions must be heavier than the electroweak scale to avoid the phenomenological constraints whereas $N$ and $N^c$ can be made very light such that it can form a condensate below the confinement scale $\sim 4\pi f_{\pi'}$. The hierarchy of $m_L \gg f_{\pi'} \gg m_N$ will be assumed as in~\cite{Graham:2015cka}.
%
%%%%%%%%%%%%%%%%%%%%%%
\begin{figure}[tph]
\begin{center}
\includegraphics[width=0.46\textwidth]{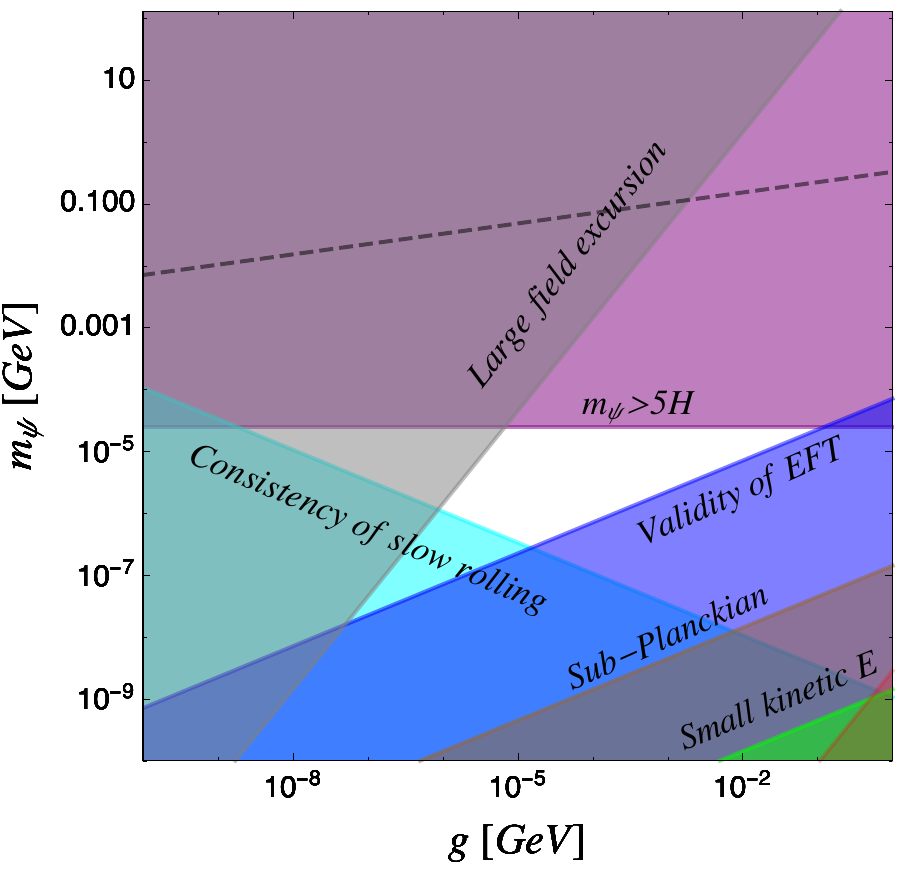}\quad
\includegraphics[width=0.445\textwidth]{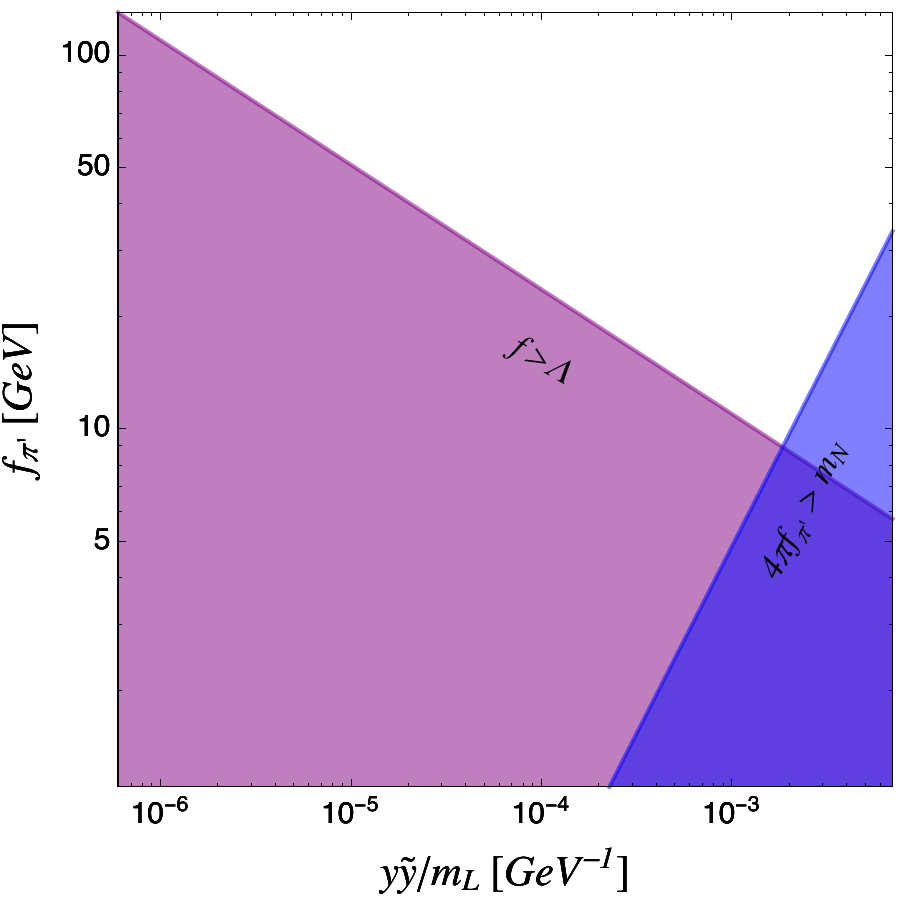}\quad
\caption{\small The viable parameter space for non-QCD model. The light-red region in bottom-right corner of the first panel corresponds to the excluded region from the small energy density carried by the dark fermion. The constraint from the `classing rolling beats the quantum spreading' is too weak to show up in the plot. We set $\Lambda = 10^4$ GeV, $H=5 \times10^{-6}$ GeV, $f_\psi = 0.5$ GeV, and $N_e=100$ in both panels (and $g = 10^{-6}$ as well for the right panel). The approximation $\mu^2 \ll \xi$ is not valid in the region above the dashed line.}
\label{fig:excl:nonQCD}
\end{center}
\end{figure}
%%%%%%%%%%%%%%%%%%%%%%

Assuming that the anomalous interaction $(\phi/f)\, G'_{\mu\nu}{\tilde G}'^{\mu\nu}$ is allowed in the model, it can be traded for the phase of the mass term $m_N$ via the chiral rotation for $N$, 
\begin{equation}\label{eq:mN:cosine}
m_N e^{i\phi/f} NN^c + {\rm h.c.} = m_N NN^c \cos\frac{\phi}{f}~, 
\end{equation}
where $m_N$ collectively refers to not only the bare mass in Eq.~(\ref{eq:nonQCD:mass:Yukawa}) but also all kinds of corrections to the mass. An analogous term to Eq.~(\ref{eq:mN:cosine}) will generate the periodic potential for $\phi$ of the type $\sim \Lambda_c^4 \cos (\phi/f)$ through the condensate $\langle NN^c \rangle \sim 4\pi f^3_{\pi'}$ below the confinement scale. While $m_N$ in Eq.~(\ref{eq:mN:cosine}) gets various contributions, the $h$-dependent contribution at tree level is estimated to be
\begin{equation}\label{eq:Lc:nonQCD}
 \Lambda^4_c = 4 \pi f^3_{\pi'}\, m_N \sim 4\pi f^3_{\pi'} \frac{y\tilde{y}\langle h \rangle^2}{m_L}~,
\end{equation}
where $f_{\pi'}$ is the chiral symmetry breaking scale of new confining gauge group. For the mechanism to work, all $h$-independent contributions to $m_N$ must be subleading~\cite{Graham:2015cka}:
\begin{equation}\label{eq:natural:nonQCD}
  f_{\pi'} < \langle h \rangle \quad {\rm and} \quad m_L < \frac{4\pi \langle h \rangle}{\sqrt{\log \Lambda/m_L}}~,
\end{equation}
and it implies that the masses of $L$ and $L^c$ are at order of a few hundred GeV (lighter masses will be constrained at the LHC).  
As was mentioned above, demanding that $N$ and $N^c$ should be lighter than the confinement scale gives rise to
\begin{equation}\label{eq:LconBiggermN:nonQCD}
   4\pi f_{\pi'} > \frac{y\tilde{y}\langle h \rangle^2}{m_L}~.
\end{equation}
Using the relation in Eq.~(\ref{eq:stopping:phi}) with the expression of $\Lambda_c$ in Eq.~(\ref{eq:Lc:nonQCD}), the decay constant in the cosine potential can be expressed as
\begin{equation}\label{eq:fpi:nonQCD}
   f = 4\pi f^3_{\pi'} \frac{y\tilde{y} \langle h \rangle^2}{m_L}\frac{1}{g\Lambda^2}~,
\end{equation}
and it needs to be bigger than the cutoff scale, $f \gtrsim \Lambda$, to be consistent with the EFT.

The viable parameter space for the relaxation from the dark fermion production to work is illustrated in $(g,\, m_\psi)$ plane in the left panel of Fig.~\ref{fig:excl:nonQCD}, while fixing other parameters such as $\Lambda$, $H$, $f_\psi$, and $N_e$. A different choice of those parameters leads to a different allowed region in $(g,\, m_\psi)$ plane. In the right panel of Fig.~\ref{fig:excl:nonQCD}, we illustrate the allowed region of the parameters (from Eqs.~(\ref{eq:LconBiggermN:nonQCD}) and~(\ref{eq:fpi:nonQCD})), which are specific to the non-QCD model, in the $(y\tilde{y}/m_L,\, f_{\pi'})$ plane for the cutoff $\Lambda = 10^4$ GeV. 
Based on the result in Fig.~\ref{fig:excl:nonQCD}, we present one benchmark point in Table~\ref{tab:benchmark:nonQCD} for the illustration. The large field excursion of $\phi$ is definitely sub-Planckian, $\Delta \phi \sim \Lambda^2/g < M_p$, for the range of $g$ in Fig.~\ref{fig:excl:nonQCD} and cutoff scale $\Lambda \sim 10^4$ GeV.

%%%%%%%%%%%%%%%%
\begin{table}
\centering
\begin{tabular}{|c|c|c|c|c|c|c|c|c|c|c|}
\hline
 $\Lambda$ & $H$ & $m_\psi$& $f_\psi$ & $g$ & $m_L$ &  $y\tilde{y}$ & $f_{\pi'}$ & $f$ & $m_\phi$ \\
 \hline 
$10^4$ &$5 \times 10^{-6}$&$1. \times 10^{-6}$ & $0.5$ & $1. \times 10^{-6}$ & $300$ & $ 1.5\times10^{-2}$ &  $45$ & $3.4\times 10^4$ & $5.\times 10^{-2}$ \\
\hline
\end{tabular} 
\caption{A benchmark point in GeV (except the dimensionless parameter $y\tilde{y}$) for the non-QCD model.
}
\label{tab:benchmark:nonQCD}
\end{table}

If the energy released from the strong fermion production can be transferred to the visible sector with the new strong group, the barrier $\Lambda_c$ will disappear during the reheating-era and the relaxion will start rolling again, spoiling the mechanism. This unwanted property has led to non-trivial constraints on the model~\cite{Graham:2015cka}. The situation is worse in our scenario with the dark fermion production as there could be a chance that the SM sector might be thermalized even during the inflation, spoiling the entire mechanism. One way to resolve these issues can be found in the so-called double scanner mechanism~\cite{Espinosa:2015eda} where the new strong gauge group is engineered to get strongly coupled at the same scale as the cutoff value. The detailed analysis of the double scanner mechanism in the context of the strong dark fermion production will be the subject of the next section.

%%%%%%%%%%%%%%%%%%%%%%%%
%%%%%%%%%%%%%%%%%%%%%%%%
\subsection{Double scanner mechanism}
\label{sec:doublescanner}	
The double scanner mechanism~\cite{Espinosa:2015eda} introduces an additional slow-rolling field $\sigma$ whose main role is controlling the amplitude of the cosine potential, while scanning the Higgs mass parameter is still carried out by the original relaxion field $\phi$. The relevant part of the potential is given by

\begin{equation}\label{eq:pot:dscanner}
\begin{split}
 \Delta V =&\  g \Lambda^2 \phi + g_\sigma \Lambda^2 \sigma + \left ( -\Lambda^2 + g \phi \right ) |h|^2 + A(\phi,\, \sigma,\, h) \cos \left (\phi/f \right )
 \\[2pt]
 &\ + \frac{\partial_\mu \phi}{f_\psi} \bar{\psi} \gamma^\mu \gamma^5 \psi + \frac{\partial_\mu \sigma}{f_\sigma} \bar{\psi} \gamma^\mu \gamma^5 \psi + m_\psi \bar{\psi}\psi~,
\end{split}
\end{equation}
where we will assume the universal decay constants $f_\psi = f_\sigma$ for $\phi$ and $\sigma$ for simplicity. The amplitude $A(\phi,\, \sigma,\, h)$ in Eq.~(\ref{eq:pot:dscanner}) is given by
\begin{equation}\label{eq:Amp:dscanner}
  A(\phi,\, \sigma,\, h) = \epsilon \Lambda^4 \left ( \beta + c_\phi \frac{g \phi}{\Lambda^2} - c_\sigma \frac{g_\sigma \sigma}{\Lambda^2}  + \frac{|h|^2}{\Lambda^2}\right )~,
\end{equation}
where $\epsilon$ is a supurion that accounts for the shift symmetry breaking.
Recall that the confinement scale in non-QCD model has to be much lower than the cutoff scale to suppress all $h$-independent contribution to the cosine potential individually (see Eq.~(\ref{eq:natural:nonQCD}) for instance). In the double scanner mechanism, the confinement scale can be made as big as the cutoff scale, $\Lambda_{\rm con} \sim \Lambda$, while keeping $h$-dependent contribution in Eq.~(\ref{eq:Amp:dscanner}) as the dominant term. The newly introduced slow-rolling field $\sigma$ cancels all $h$-independent terms of order $\sim \epsilon \Lambda^4$ in Eq.~(\ref{eq:Amp:dscanner}) together during the cosmological evolution of interest with the appropriate choice of coefficients, $\beta$, $c_\phi$, and $c_\sigma$ of order one, and this cancellation is technically natural.

The double scanner mechanism features multiple stages of cosmological evolution in $(\phi,\, \sigma)$ plane. In stage one, two fields $\phi$ and $\sigma$ start evolving at some initial points while $\phi\gtrsim \Lambda/g$ and $\sigma\gtrsim\Lambda/g_\sigma$ such that the effective Higgs mass squared parameter is positive. The electroweak symmetry is unbroken as usual. Since the amplitude has a generic size of $A \sim \epsilon \Lambda^4$, the potential for $\phi$ is dominated by the $A \cos \left (\phi/f \right )$ which causes $\phi$ to get stuck at some minimum. The $\sigma$ field continuously rolls down while scanning the amplitude $A$. At some point, the cosine potential for $\phi$ becomes smaller than the linear potential for $\phi$, and $\phi$ starts rolling down the linear potential (the second stage begins). The field $\phi$ continues rolling down along the trajectory, represented by $\phi_*$ in~\cite{Espinosa:2015eda}, along which the evolution is dominantly driven by the linear potential. The second (third) stage ends (begins) when $\phi$ passes the critical point, $\phi_c = \alpha \Lambda/g$, after which the sign of the effective Higgs mass squared flips, and the electroweak symmetry breaking is triggered. In stage three, the $h^2$ term in Eq.~(\ref{eq:Amp:dscanner}) is switched on, and it dominates the amplitude. The amplitude scaling as $\sim h^2$ keeps growing as $\phi$ keeps rolling down the potential since the negative Higgs mass squared term increases. The relaxion field stops rolling and it is trapped in one of the cosine potential wells when the steepness of the cosine potential is balanced with the slope of the linear potential: 
\begin{equation}\label{eq:dscanner:stopping}
  g\Lambda^2 =\frac{A}{f} \sim \frac{\epsilon \Lambda^2 v^2}{f}.
\end{equation}
In the last stage, the field $\sigma$ keeps moving down the potential~\footnote{While there is a trajectory along which the amplitude is the smallest in presence of the positive $h^2$ term, evolving along that trajectory requires the negative velocity of $\sigma$ (since $\sigma$ should climb up the potential in the $\sigma$ direction to increase the $\sigma$ value).} until it finds its minimum somewhere. Around the end of the last stage, we do not expect any cancellation in the amplitude $A(\phi,\, \sigma,\, h)$, and the typical size of the amplitude will be $A \sim \epsilon \Lambda^4$ which implies that the natural size of $\phi$ mass without fine-tuning is expected to be
\begin{equation}\label{eq:mphi:dscanner}
 m_\phi^2 = \frac{\epsilon \Lambda^4}{f^2} \sim \frac{g}{v^2}\frac{\Lambda^4}{f} = \frac{g}{f}\left ( \frac{\Lambda}{v} \right )^4 v^2~.
\end{equation}
Whereas the mass of $\sigma$ is given by
\begin{equation}
  m^2_\sigma \sim g_\sigma^2~.
\end{equation}
For the successful scanning over $\phi$ tracking $\sigma$ along the sliding trajectory $\phi_*$ (where the cosine potential is smaller than the linear potential) before it reaches the critical point, $\phi_c$, the condition $d\phi(t)/d\sigma(t) = (g/g_\sigma)^{1/2} > d\phi_*/d\sigma$ should be satisfied~\footnote{The slow roll velocities, $\dot\phi$ and $\dot\sigma$, in our scenario using fermion production are determined by the nonlinear backreaction terms, and it leads to $d\phi(t)/d\sigma(t) = (g/g_\sigma)^{1/2}$ instead of $g/g_\sigma$ (=$V_\phi/V_\sigma$) as in~\cite{Espinosa:2015eda}.}. Otherwise, $\phi$ gets deviated from the trajectory before it reaches $\phi_c$, and it gets stuck at some minimum. To avoid this situation, we require
\begin{equation}
  c_\phi\, g^{3/2} > c_\sigma\, g^{3/2}_\sigma~.
\end{equation}
which gives rise to $g > g_\sigma$ for $c_\phi \sim c_\sigma \sim \mathcal{O}(1)$. Once it passes the critical point where the Higgs barrier is switched on, $\phi$ needs to exit the trajectory to continue its evolution along the path where the amplitude grows like $\sim \epsilon \Lambda^2 v^2$. It requires $d\phi(t)/d\sigma(t) < d\phi_*/d\sigma$ which leads to $(c_\phi - 1/(2\lambda) )\, g^{3/2} > c_\sigma\, g^{3/2}_\sigma$ (see~\cite{Espinosa:2015eda} for the related discussion).

Theoretical constraints for the double scanner mechanism to work are similar to those in non-QCD model in Section~\ref{sec:nonQCD}. The constraints from Eq.~(\ref{eq:constraints:EFT}) to Eq.~(\ref{eq:classrolling:constraint}) similarly apply to $\sigma$ with $g_\sigma$, and we take a stronger one in the numerical simulation to determine the viable parameter space. Since the cosine potential for $\phi$ contributes to the Higgs mass squared of order $\sim m^2_\phi$, we demand $m^2_\phi \lesssim v^2$ not to reintroduce the naturalness problem. Using the expression in Eq.~(\ref{eq:mphi:dscanner}), we obtain the upper bound on $g$,
\begin{equation}\label{eq:dscanner:g1}
  \frac{g}{f} \left ( \frac{\Lambda}{v} \right )^4 \lesssim 1 \quad \rightarrow \quad g \lesssim f \left ( \frac{v}{\Lambda} \right )^4~.
\end{equation}
While the double scanner mechanism relies on the field $\sigma$ which cancels the amplitude in Eq.~(\ref{eq:Amp:dscanner}), the cosine potential gets quantum corrections at quadratic order on $\cos (\phi/f)$ such as $\epsilon^2\Lambda^4 \cos^2 (\phi/f)$ etc~\cite{Espinosa:2015eda}. The mechanism is spoiled unless those corrections to Higgs barrier ($\sim \epsilon \Lambda^2 v^2$) remain subleading, and ignoring the loop factors, it requires
\begin{equation}\label{eq:dscanner:epsilon}
 \epsilon \lesssim v^2/\Lambda^2~.
\end{equation}
Combing Eq.~(\ref{eq:dscanner:stopping}) and Eq.~(\ref{eq:dscanner:epsilon}), we obtain another upper bound on $g$,
\begin{equation}\label{eq:bound:g}
  g \lesssim \frac{v^4}{f \Lambda^2} = \left (\Lambda/f \right )^2  f \left ( \frac{v}{\Lambda} \right )^4 \lesssim \frac{v^4}{\Lambda^3}~, 
\end{equation}
where the last inequality is due to $f \gtrsim \Lambda$ which also makes the above constraint stronger than Eq.~(\ref{eq:dscanner:g1}). 

%%%%%%%%%%%%%%%%%%%%%%
\begin{figure}[tph]
\begin{center}
\includegraphics[width=0.46\textwidth]{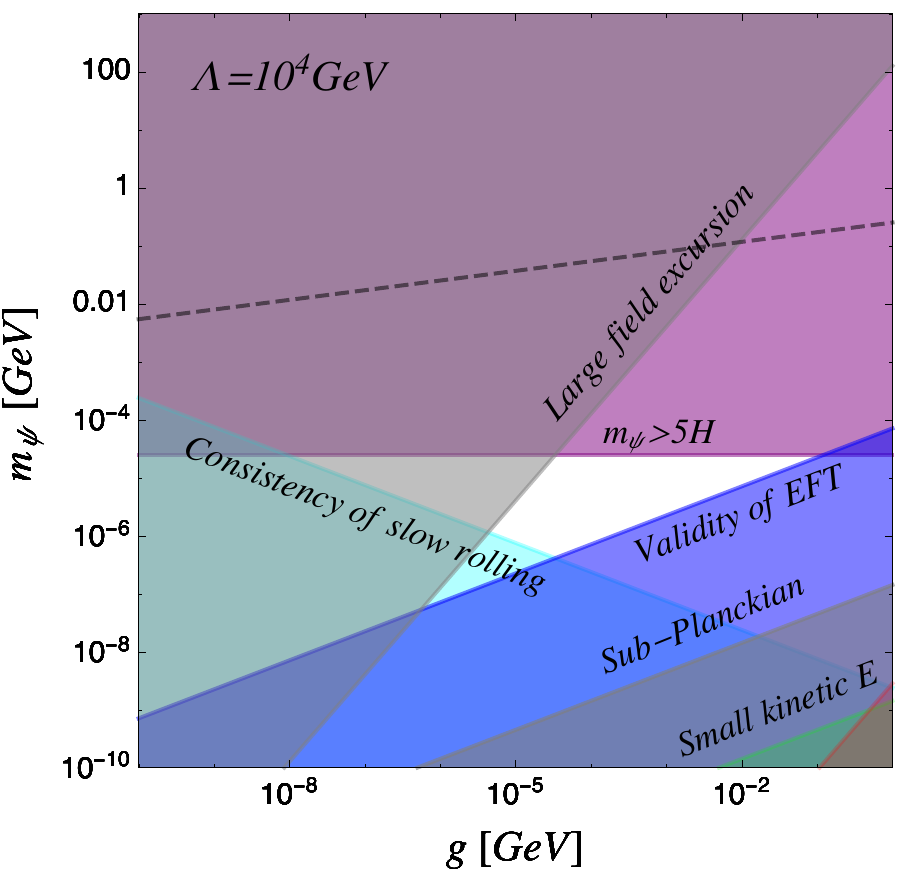}\quad
\includegraphics[width=0.462\textwidth]{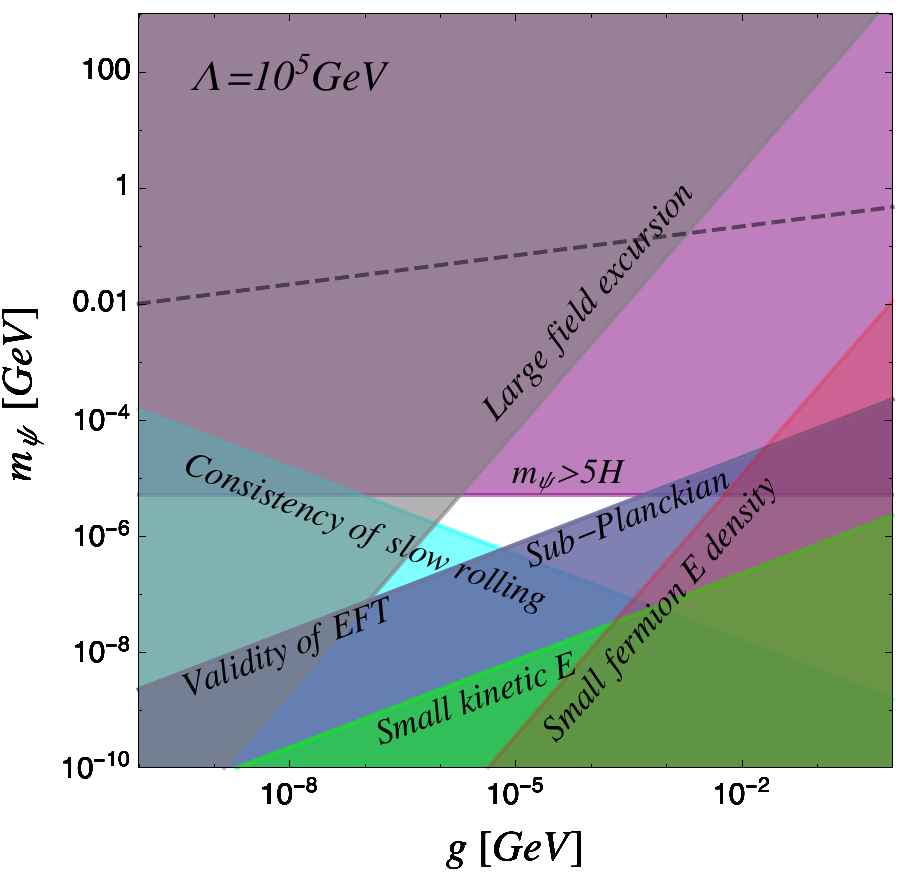}
\caption{\small The viable parameter space for the double scanner mechanism. Excluded regions from `Sub-Planckian' and `Validity of EFT' numerically coincide. The constraint from the `classical rolling beats the quantum spreading' is too weak to show up in the plot. We set $\Lambda = 10^4$ ($10^5$) GeV, $H=5 \times 10^{-6}$ ($1 \times 10^{-6}$) GeV, $f_\psi = 0.5$ (5) GeV, $\epsilon = 1\times 10^{-5}$ ($2 \times10^{-6}$), $N_e=100$, and $g_\sigma = 0.2\, g$ in the left (right) panel. The approximation $\mu^2 \ll \xi$ is not valid in the region above the dashed line.}
\label{fig:excl:doubleScanner}
\end{center}
\end{figure}
%%%%%%%%%%%%%%%%%%%%%%

%
The parameter space consistent with all the constraints mentioned above in $(g,\, m_\psi)$ plane is illustrated in Fig.~\ref{fig:excl:doubleScanner}, and a benchmark point is presented in Table~\ref{tab:benchmark:ds}. The cutoff scales were fixed to $\Lambda = 10^4$ or $10^5$ GeV in Fig.~\ref{fig:excl:doubleScanner} as our aim is to address  the little hierarchy problem (see caption of Fig.~\ref{fig:excl:doubleScanner} for other chosen parameters). 
%%%%%%%%%%%%%%%%
\begin{table}[tph]
\centering
\begin{tabular}{|c|c|c|c|c|c|c|c|c|c|}
\hline
 $\Lambda$ & $H$ & $m_\psi$& $f_\psi \sim f_\sigma$ & $g$ & $g_\sigma (\sim m_\sigma)$ &  $\epsilon$ & $f$ & $m_\phi$ \\
 \hline 
$10^4$ &$5 \times 10^{-6}$&$1. \times 10^{-6}$ & $0.5$ & $1. \times 10^{-5}$ & $2. \times 10^{-6}$ & $ 1.\times10^{-5}$ &  $6.1\times 10^4$ & $5.2$ \\
 \hline
$10^5$ &$1 \times 10^{-6}$&$1. \times 10^{-6}$ & $5$ & $1. \times 10^{-6}$ & $2. \times 10^{-7}$ & $2.\times 10^{-6}$ & $1.2\times 10^{5}$ & $1.2 \times 10^2$ \\
\hline
\end{tabular} 
\caption{A benchmark point in GeV (except the dimensionless parameter $\epsilon$) for the double scanner mechanism.
}
\label{tab:benchmark:ds}
\end{table}

As was explained in Section~\ref{sec:theory:constraint}, the inflation is driven by a separate inflaton sector to avoid scanning the thermal Higgs mass squared during the scanning era. This idea works for the double scanner mechanism due to the confinement scale as big as the cutoff scale unlike the case of the non-QCD model. However, the successful double scanner mechanism from the fermion production also has to be safe against being thermalized to the temperature above $\Lambda$ during the reheating era, and it imposes a constraint on the inflaton sector. In this work, without getting into the details of the reheating, we assume that the energy stored in the inflaton sector is smaller or at most comparable to that of the relaxation sector (collectively denotes the entire sector of $\psi$, $\phi$, SM, and the strong gauge group) when the relaxation sector gets in thermal equilibrium~\footnote{A possibility to make the inflaton sector energy density to be subdominant is to require that the inflaton decays into radiation after inflation and that this timescale is sufficiently short compared to the thermalization timescale of the relaxation sector (see~\cite{Adshead:2019aa} for a recent discussion about reheating in different sectors). In a situation that the reheating temperature in the inflaton sector is higher than the cutoff scale $\Lambda$ and the relaxation sector is in thermal equilibrium with the inflaton sector during such a short time scale, thermal effects would erase the periodic potential barriers, leading to the stabilized relaxion to roll again. The inflaton sector and its interaction to the relaxation sector have to be constrained such that either the second scanning era itself does not happen or it does not overshoot the electroweak scale during the second scanning era (if it occurs). Another possibility is to consider a scenario where its energy density decays faster than the radiation~\cite{Liddle:2003as,Kobayashi:2010cm}.}.  We also assume that the end of relaxation coincides with the end of inflation such that the fermion energy density $\rho_\psi$ would not be diluted away by inflation to allow the possibility of $\psi$ as a dark matter candidate. This can be achieved by appropriately setting up the inflationary sector and choosing parameters therein.

%%%%%%%%%%%%%%%%%%%%%%%%
\subsection{Comments on $f_\psi \ll \Lambda$}
\label{sec:comment:ther:decoup}

Apparently, the scale $f_\psi$ in the derivative interaction in Eq.~(\ref{eq:deri:coupling}) needs to be much smaller than the cutoff scale of the model, namely $f_\psi \ll \Lambda$, to make the fermion production efficient enough. Two hierarchical scales without a natural explanation can be considered to be either inconsistent from the EFT point of view or a strong coupling problem when  $f_\psi$ is normalized to $\Lambda$. It has been a generic issue in applications which heavily rely on the fermion production as a main dissipation, and we are not an exception. While we do not have a solution for this issue, we will briefly comment on the possibility of the thermal decoupling between the dark fermion sector and the visible sector with the new strong gauge group.

The thermal decoupling is an attractive idea, if it can be realized, for two theoretical issues: the second scanning in the reheating era and two apparently inconsistent scales within an EFT. If the interaction rate between two sectors can be made negligible over the entire range of temperature with respect to the Hubble parameter, the dark fermion sector will have its own thermal history without interfering with the visible sector. If this is the situation, the barrier of the cosine potential will not be erased.  Besides, 
it will be natural for two thermally disconnected sectors to have their own cutoff scales, and the EFT of one sector would not spoil the validity of the EFT of the other sector. 

%%%%%%%%%%%%%%%%%%%%%%
\begin{figure}[tph]
\begin{center}
\includegraphics[width=0.46\textwidth]{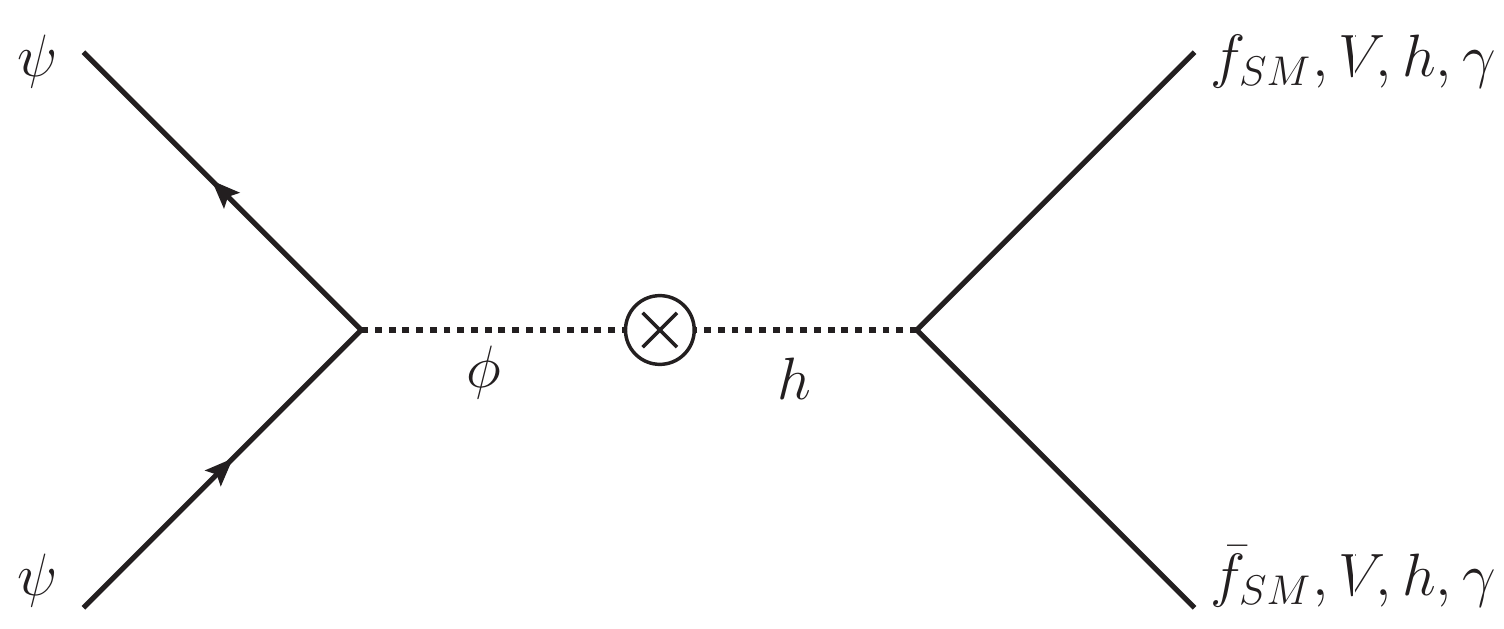}\quad
\includegraphics[width=0.27\textwidth]{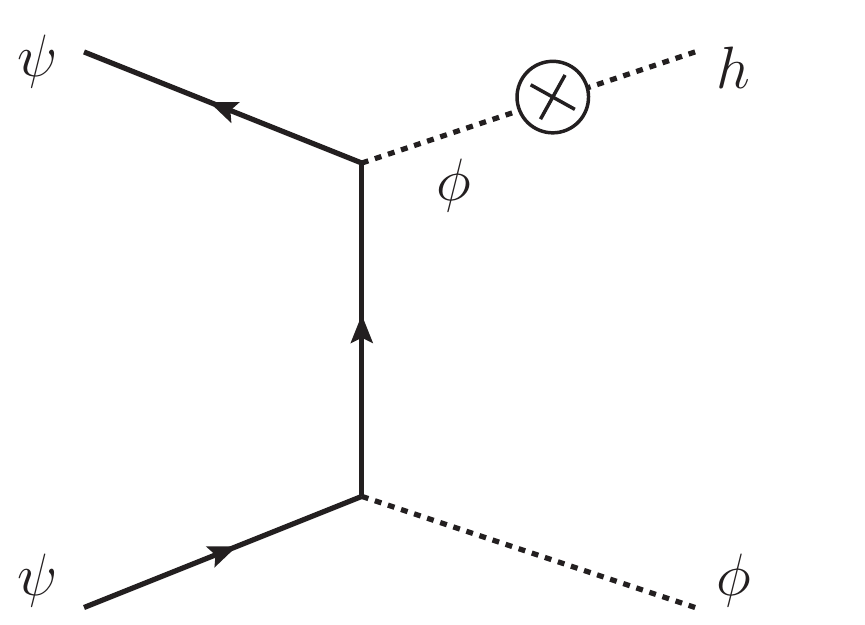}
\caption{\small  Typical diagrams for the interaction between the dark fermion sector and SM sector where at least one $\phi$ appears as an intermediate mediator.}
\label{fig:decoupling:case1}
\end{center}
\end{figure}
%%%%%%%%%%%%%%%%%%%%%%

From our numerical simulation, we find that the interaction rate between two sectors via the intemediate $\phi$-exchange such as diagrams in Fig.~\ref{fig:decoupling:case1} can be made smaller than the Hubble parameter. It is basically because the diagrams are doubly suppressed: small $m_\psi$ parameter from the derivative coupling and $h$-$\phi$ mixing angle. There also could be processes that decay into the states of the new strong gauge group. Based on our naive dimensional analysis (NDA), we find that the interaction rate can be made smaller for the double scanner scenario (marginally smaller for the non-QCD model) than the Hubble parameter although it has to be confirmed by more exact numerical simulation.

%%%%%%%%%%%%%%%%%%%%%%
\begin{figure}[tph]
\begin{center}
\includegraphics[width=0.29\textwidth]{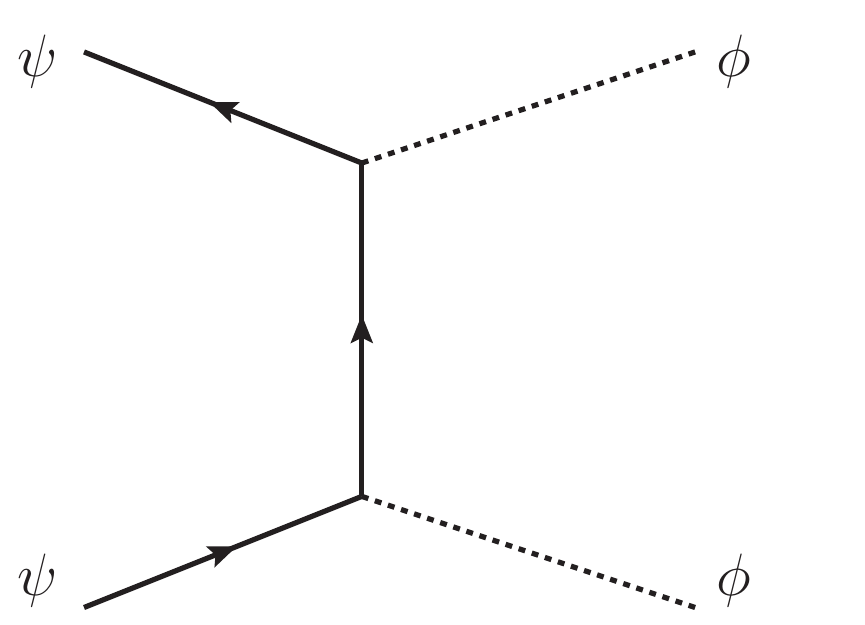}\quad
\includegraphics[width=0.29\textwidth]{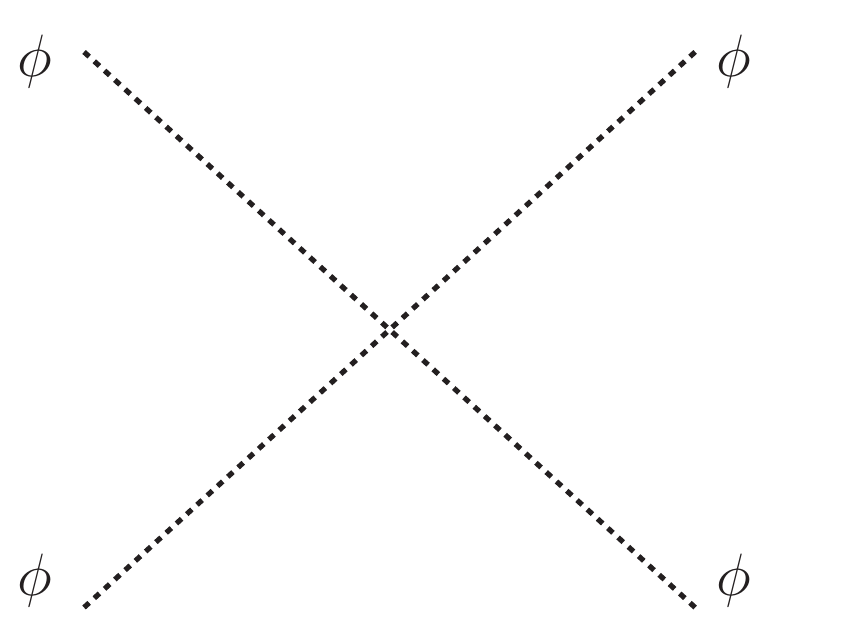}\quad
\includegraphics[width=0.29\textwidth]{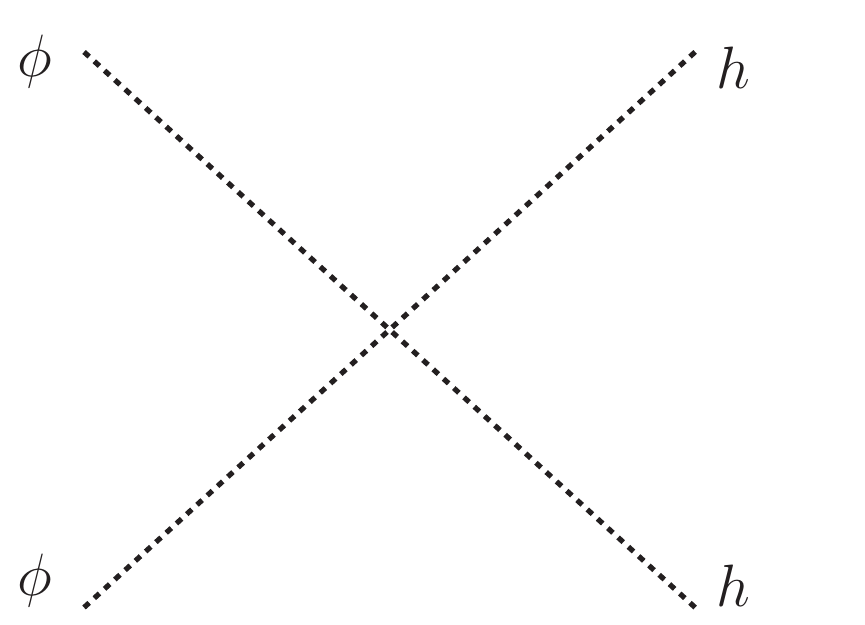}
\caption{\small Example diagrams for three different types of interactions between 1) dark fermion sector and $\phi$ sector, 2) self interaction in the $\phi$ sector, and 3) $\phi$ sector and SM sector.}
\label{fig:decoupling:case2}
\end{center}
\end{figure}
%%%%%%%%%%%%%%%%%%%%%%

The major threats to the idea of the thermal decoupling come from the processes such as those in Fig.~\ref{fig:decoupling:case2}. For instance, the dark fermion can first thermally produce the relaxion $\phi$ whose lifetime is not shorter than the typical interaction timescale. Then, those $\phi$ fields can thermally produce the SM particles. The typical diagrams such as those in Fig.~\ref{fig:decoupling:case2} have only one suppression which is not enough to make the interaction rate smaller than the Hubble parameter~\footnote{As was mentioned in Section~\ref{sec:theory:constraint}, the situation is different during the inflation as the inflationary expansion rate $H$ during the inflation is assumed to be higher than $\Lambda^2/M_p$. }. In this situation, the energy carried by the dark fermion will be efficiently transferred to the visible sector. On the other hand, when the temperature drops below $m_\phi$, the diagrams involving $\phi$ in Fig.~\ref{fig:decoupling:case2} will be exponentially suppressed as $\phi$ becomes non-relativisitc particle, and thus two sectors will be thermally decoupled. 

The above discussion suggests that one need to suppress those diagrams in Fig.~\ref{fig:decoupling:case2} to achieve the thermal decoupling over the entire temperature range. One might try to dilute the energy density released by the dark fermion away below $m_\phi$ such that the diagrams in Fig.~\ref{fig:decoupling:case2} are exponentially suppressed. 
%We have not managed to resolve the issue, and 
We leave it for the future work.

%%%%%%%%%%%%%%%%%%%%%%%%
%%%%%%%%%%%%%%%%%%%%%%%%
%%%%%%%%%%%%%%%%%%%%%%%%
\section{Prospect for dark matter}
\label{sec:DM}
We investigate the compatibility of two models in Sections~\ref{sec:nonQCD} and~\ref{sec:doublescanner} with the current phenomenological and astrophysical constraints.

\subsection{Relic abundance and dark matter}
Since the Higgs field dominantly mixes with $\phi$, we diagonalize only the $2\times 2$ mass matrix of $(h,\, \phi)$, and we treat $\sigma$ separately.
\begin{equation}
\begin{pmatrix} h_1 \\ h_2 \end{pmatrix} = \begin{pmatrix} \cos\theta & \sin\theta \\ - \sin\theta & \cos\theta \end{pmatrix} \begin{pmatrix} h \\ \phi \end{pmatrix}~,
\end{equation}
where the mixing angle is given by
\begin{equation}
  \tan (2\theta) = \frac{2\, m_{h\phi}}{m^2_h - m^2_\phi}~,\quad {\rm or}\quad \tan\theta = \frac{y}{1+\sqrt{1+y^2}}~,
\end{equation}
where we defined $y\equiv 2\, m_{h\phi}/(m^2_h - m^2_\phi)$, and $m_{h\phi} = gv$. Since $m^2_h > m^2_\phi$ is required not to reintroduce the fine-tuning, the mixing angle is roughly $\sim 2\, m_{h\phi}/m^2_h \sim 2 gv/m_h^2$.

The decay rate of $\phi$ in non-diagonalized basis is given by
\begin{equation}
 \Gamma_\phi= \theta^2_{\phi h}\Gamma_h (m_\phi ) + \Gamma_{\phi\rightarrow \psi\psi} (m_\phi)~,
\end{equation}
where 
\begin{equation}
\Gamma_{\phi\rightarrow \psi\psi} 
%= \frac{1}{2 m_\phi} \frac{8 m_\psi^2\, m_\phi^2}{f_\psi^2} \frac{1}{8\pi} \sqrt{1 - \frac{4m_\psi^2}{m_\phi^2}} 
= \frac{1}{2\pi} \frac{m_\psi^2}{f_\psi^2} m_\phi \sqrt{1 - \frac{4m_\psi^2}{m_\phi^2}}~.
\end{equation}
It is important to notice that the derivative coupling gives an effective coupling of $m_\psi/f_\psi$ instead of $E$-growing effective coupling $E/f_\psi$. One can see that the suppression by $m_\psi/E$ is originated from the equation of motion of fermions in the derivative coupling. The decay rate for $\sigma$ is obtained by the replacement $\phi \leftrightarrow \sigma$. While $\phi$ can always decay into dark fermions before the Big Bang Nucleosynthesis (BBN), the $\sigma \rightarrow \psi\psi$ channel might not be kinematically allowed in some allowed parameter space where $m_\sigma \sim g_\sigma < 2\, m_\psi$. 
Once the $\sigma \rightarrow \psi\psi$ decay channel is kinematically opened, the $\sigma$ field decays into dark fermions well before BBN. Otherwise, $\sigma$ mainly decays into SM fields via the mixing with the Higgs, and its lifetime is longer than the age of the universe. Therefore, the candidates that could potentially serve as dark matters in our scenario are $\psi$ and $\sigma$.

The abundance of $\sigma$ can get a contribution from the vacuum misalignment. The energy density $\rho_\sigma$ at the start of the oscillating regime is $\sim m^2_\sigma (\Delta \sigma )^2$ where the misalignment of $\sigma$ is given by $\Delta\sigma \sim \sqrt{N_e} H$~\footnote{This estimate of the misalignment is better justified when $m_{\sigma}\ll H$ (with $H$ being the Hubble during the inflation). While $m_\sigma \lesssim H$ in our benchmark points, our conclusion on the abundance of $\sigma$ remains the same. Also note that the quantum fluctuations for $m_\sigma > H$ are further suppressed~\cite{Birrell:1982aa}.}. The relic abundance for $N_e\sim {\mathcal{O}(10^2)}$ is expected to be negligible,
\begin{equation}\label{eq:abundance:sigma}
  \Omega^\sigma_0 = \frac{\rho^\sigma_0}{\rho_c} \sim \frac{1}{\rho_c} m^2_\sigma N_e H^2  \left (\frac{T_0}{\sqrt{m_\sigma M_p}} \right )^3 \ll 1~,
  \end{equation}
where $T_{osc}=\sqrt{m_\sigma M_p}$ is the temperature below which $\sigma$ is in the oscillating regime. We conclude that the non-thermally produced $\sigma$ has negligible contribution to the current dark matter abundance.
%%%%%%%%%%%%%%%%%%%%%%
\begin{figure}[tph]
\begin{center}
\includegraphics[width=0.46\textwidth]{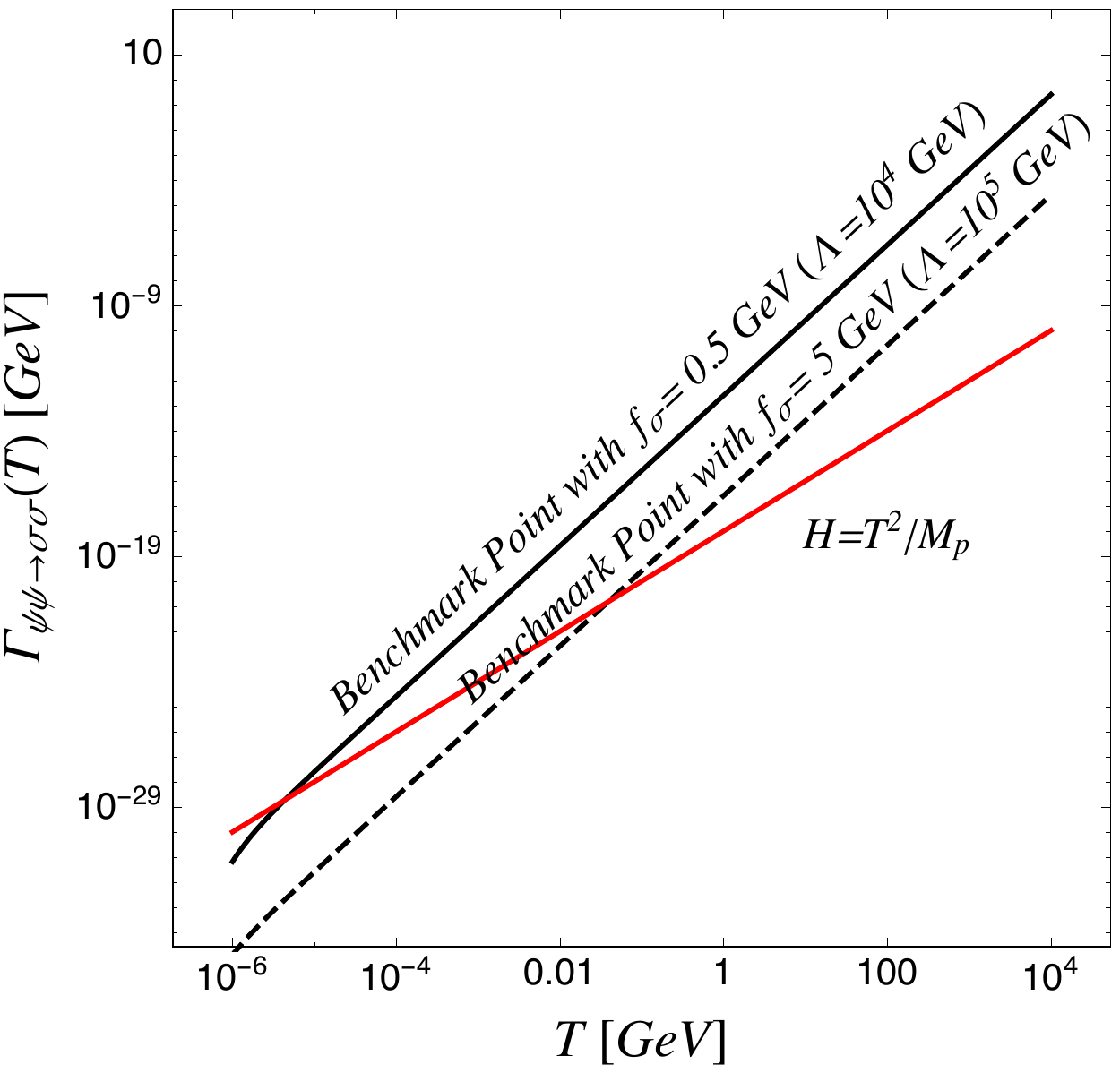}\quad
\caption{\small The thermally averaged interaction rate of $\psi\psi \rightarrow \sigma\sigma$, $\Gamma_{\psi\psi\rightarrow\sigma\sigma}(T)$, (black) for two benchmark points in Table~\ref{tab:benchmark:ds} and the Hubble rate (red).}
\label{fig:decoupling:psi2sigma}
\end{center}
\end{figure}
%%%%%%%%%%%%%%%%%%%%%%

The $\sigma$ field can also be produced from the thermal bath via $\psi\psi \rightarrow \sigma\sigma$ (via $t$ and $u$-channels)  process. The production of $\sigma$ from the $\phi$ scattering is negligible.
The order of magnitude estimate of the thermally averaged interaction rate, ignoring all numeric factors, in the limit $T \gg m_\psi, \, m_\sigma$ is roughly
\begin{equation}
  \Gamma_{\psi\psi\rightarrow \sigma\sigma}(T) \sim \frac{m^2_\psi}{f^4_\sigma} T^3~.
\end{equation}
The thermally produced $\sigma$ field is expected to decouple from the thermal bath below the temperature,
\begin{equation}\label{eq:decoup:psi2sigma:NDA}
  T_d \sim \frac{f^4_\sigma}{M_p} \frac{1}{m^2_\psi}~,
\end{equation}
which looks close to $m_\psi$ and $m_\sigma$ especially for the case of our benchmark point with $\Lambda \sim 10^4$ GeV in Table~\ref{tab:benchmark:ds}. The more exact numerical evaluation of the thermally averaged interaction rate is illustrated in Fig.~\ref{fig:decoupling:psi2sigma} for two benchmark points in Table~\ref{tab:benchmark:ds}, and the decoupling temperatures are found to be roughly two orders of magnitudes higher than our rough estimate in Eq.~(\ref{eq:decoup:psi2sigma:NDA}). It implies that the thermally produced $\sigma$ gets out of equilibrium while being relativistic. Among the annihilation channels of $\psi$ into a pair of $\sigma$, $\phi$, or SM particles, the dominant channels are $\psi\psi\rightarrow \phi \phi,\, \sigma\sigma$ (the channel to $\phi \phi$ will be shut off below $m_\phi$ though). Similarly $\psi$ decouples from the thermal bath while being relativistic.

Both $\psi$ and $\sigma$ (or only $\psi$ if $\sigma$ decays before BBN) could be warm dark matter (WDM) candidates, and their abundances today are, assuming no entropy dilution with the constant relativistic degrees of freedom from its decoupling epoch, 
  \begin{equation}\label{eq:relic:psi:sigma1}
   \Omega_0^{\sigma,\, \psi} = \frac{m_{\sigma,\, \psi}\, n_{\gamma}^0}{\rho_c} \sim 
%   \frac{m_{\sigma,\, \psi}}{0.025 \text{keV}\, h^2},
 \frac{0.12}{h^2}  \left ( \frac{m_{\sigma,\, \psi}}{3.\times 10^{-3} \text{keV}} \right ),
\end{equation}
where we have used the fact that after decoupling the photon temperature $T_\gamma$ and the effective temperature of $\sigma,\,\psi$ evolve identically due to the constant relativistic degrees of freedom, namely $T_\gamma = T_{\sigma,\,\psi}$. For our keV-scale $\sigma$ and $\psi$, the relic abundance in Eq. (\ref{eq:relic:psi:sigma1}) is three orders of magnitude higher than what is needed to be consistent with the present relic abundance. This constraint is in a tension with those from the Lyman-$\alpha$ forest analysis on the free-streaming length as they put a strong lower bound on the thermal WDM mass $m_{WDM}\gtrsim 5 \text{keV}$~\cite{Bond:1980aa,Kolb:1990aa,irsic2017}. If the warm dark matter is a partial component of the whole dark matter, the Lyman-$\alpha$ bound can be relaxed \cite{Boyarsky:2009aa}.
      
The tension between the relic abundance (\ref{eq:relic:psi:sigma1}) and the Lyman-$\alpha$ bound in thermal WDM can be ameliorated in our scenario as follows.
For $\sigma$, the easiest solution will be to make it heavier than $2m_\psi$ so that the channel $\sigma \rightarrow \psi\psi$ is kinematically opened and $\sigma$ decays into dark fermions before the BBN. This option will be viable for the first benchmark point with the cutoff, $\Lambda\sim 10^4$ GeV, in Table~\ref{tab:benchmark:ds} after a slight modification of parameters, but it will be difficult for the second benchmark point with the higher cutoff.  For $\psi$ (and $\sigma$ if the decay to $\psi$ is forbidden), for instance, one can modify the strength of $\psi\psi \rightarrow \sigma\sigma$ to significantly increase the decoupling temperature such that the discrepancy of the effective degrees of freedom at the decoupling temperature $T_d$ and at $T_0$ leads to (up to) two-order of magnitude suppression. It can be achieved by adapting non-universal fermion couplings to $\phi$ and $\sigma$ such as setting $f_\sigma = \Lambda$ while keeping $f_\psi$ as before. In this situation, the slow-rolling of $\sigma$ will be maintained via the Hubble friction~\footnote{This does not reintroduce the downsides of the GKR scenario such as the super-Planckian field excursion and a large number of e-folding.} while the fermion production is still responsible for the slow-rolling of $\phi$. This choice makes the interaction rate $\Gamma_{\psi\psi\rightarrow \sigma\sigma}(T)$ always smaller than $H$ within the cutoff scale, and thus the decoupling temperature will be roughly $m_\phi$ below which $\Gamma_{\psi\psi\rightarrow \phi\phi}$ is switched off. Using the total entropy conservation of the Universe, one can estimate $g_{*S}(T_0)/g_{*S}(T_d) \sim \mathcal{O}(10^{-2})$ for $T_d \sim m_\phi \sim \mathcal{O}(10^{1-2})$ GeV.
 Consequently, the thermal WDM mass 
bound from the Lyman-$\alpha$ constraints gets slightly relaxed due to the change in the relativistic degrees of freedom at the decoupling temperature,
\begin{equation}
    m_{WDM}\gtrsim 5 \text{keV}\times \left(\frac{g_{*S}(T_d)\sim \mathcal O(10^2)}{g_{*S}(T\ll \text{MeV})}\right)^{-1/3}\sim 1\text{keV}.
\end{equation}
Although predicted by our relaxation parameter space, the keV-scale dark fermion mass
is better consistent with this bound.  Similarly, the relic abundance of $\psi$ is diluted as
\begin{equation}\label{eq:relic:psi:sigma}
\Omega_0^{\psi} \sim 
%\frac{m_{\psi}\, n_\gamma^0}{\rho_c} \frac{g_{*S}(T_0)}{g_{*S}(T_d)} ~.
\frac{0.12}{h^2}  \left ( \frac{m_{\sigma,\, \psi}}{3.\times 10^{-3} \text{keV}} \right )  
\times \frac{g_{*S}(T_0)}{g_{*S}(T_d)} \times \frac{1}{S} ~,
\end{equation}
where we have introduced an extra suppression factor $1/S$ from the entropy dilution that might be originated from a short-period entropy injection right after the decoupling of $ \psi$. It can help lowering the relic abundance to the acceptable level~\cite{Baltz:2003aa}.
Since the interaction rates between $\sigma$ and $\phi$ with $f_\sigma = \Lambda$ are negligible compared to $H$, the $\sigma$ fields are likely non-thermal with a negligible abundance.
 
Before concluding this section, we briefly point out one significant difference between our fermion warm dark matter and the standard sterile neutrino warm dark matter. The dark fermion in our model is not only a SM singlet, but also a stable particle. The sterile neutrino, however, can decay.
Apart from decaying into three left-handed neutrinos when the sterile neutrino mass is lower than twice of the electron mass, the sterile neutrino has the loop-level decay channel: $N\to \nu+\gamma$~\cite{Pal:1982aa}, which could be detectable from the X-ray observations \cite{Abazajian:2001aa,Dolgov:2002aa}.% if the sterile neutrino is the main source of DM. 

%%%%%%%%%%%%%%%%%%%%%%%%
%%%%%%%%%%%%%%%%%%%%%%%% 
\section{Conclusion}
\label{sec:con}

We have investigated a scenario of the cosmological relaxation of the weak scale supported by the backreaction from the dark fermion production during inflation, with the cutoff scale $10^{4\sim 5}$ GeV. Being a more efficient friction source than the Hubble expansion, the fermion production plays a significant role in removing downsides in the original GKR scenario. Hence, our models do not have any extremely small parameters, the number of e-folds is of appropriate size, and the relaxion field excursion is sub-Planckian. 

A characteristic property of the models with the fermion production through the derivative coupling is the possible thermalization between the produced fermions and the axionic source field even during inflation. While this could be alarming in a relaxion model in which scanning over the thermal Higgs mass squared should be avoided, the double scanner scenario can survive (while the non-QCD model might not) by considering the relaxation during the inflation whose inflationary expansion rate is higher than $\Lambda^2/M_p$ such that the thermal relaxion cannot thermalize the visible sector during inflation.  Another generic feature, or unwanted downside, of those models lies in the appearance of a scale $f_\psi$, associated with the axionic derivative coupling to the fermion, much smaller than the EFT cutoff scale. This may lead to strong coupling, non-perturbativity, or the EFT inconsistency problem, and should be solved separately by an extra mechanism. We leave it to the future work.

Our model allows the possibility of the observation-consistent keV scale warm dark matter. While the fermion mass parameter controls the strength of the interaction in various places in our model, its preferred value for our scenario to work is coincidently in the vicinity of the right ballpark for the keV scale warm dark matter.

%%%%%%%%%%%%%%%%%%%%%%%%%%%%%%%%%%%%%%%%
%%%%%%%%%%%%%%%%%%%%%%%%%%%%%%%%%%%%%%%%
%%%%%%%%%%%%%%%%%%%%%%%%%%%%%%%%%%%%%%%%
%%%%%%%%%%%%%%%%%%%%%%%%%%%%%%%%%%%%%%%%
%%%%%%%%%%%%%%%%%%%%%%%%%%%%%%%%%%%%%%%%
%%%%%%%%%%%%%%%%%%%%%%%%%%%%%%%%%%%%%%%%
%%%%%%%%%%%%%%%%%%%%%%%%%%%%%%%%%%%%%%%%
% change__________________________________________
\section*{Acknowledgments}
%\acknowledgments
%__________________________________________
We would like to thank Pedro Schwaller and Wan-il Park for useful discussions. MU was supported by Samsung Science and Technology Foundation under Project Number SSTF-BA1602-04. MS and FY were supported by National Research Foundation of Korea under Grant Number 2018R1A2B6007000. KK was supported by Institute for Basic Science (IBS-R018-D1).  MS thanks the Aspen Center for Physics (where part of the work was done) under NSF grant PHY-1607611 for hospitality

%%%%%%%%%%%%%%%%%%%%%%%%%%%%%%%%%%%%%%%%
%%%%%%%%%%%%%%%%%%%%%%%%%%%%%%%%%%%%%%%%
%%%%%%%%%%%%%%%%%%%%%%%%%%%%%%%%%%%%%%%%
%%%%%%%%%%%%%%%%%%%%%%%%%%%%%%%%%%%%%%%%
%%%%%%%%%%%%%%%%%%%%%%%%%%%%%%%%%%%%%%%%
%%%%%%%%%%%%%%%%%%%%%%%%%%%%%%%%%%%%%%%%
%%%%%%%%%%%%%%%%%%%%%%%%%%%%%%%%%%%%%%%%
\appendix

%%%%%%%%%%%%%%%%%%%%%%%%
%%%%%%%%%%%%%%%%%%%%%%%%
%%%%%%%%%%%%%%%%%%%%%%%%
\section{Fermion production}
\label{app:FP}

%%%%%%%%%%%%%%%%%%%
%%%%%%%%%%%%%%%%%%%
\subsection{Convention}
Our convention for the explicit computations is the same as those in~\cite{Adshead:2018oaa}. The metric has mostly negative signs, $\eta^{\mu\nu} = {\rm diag.}(+1,\, -1,\, -1,\, -1)$. The gamma matrices are given by
\begin{equation}
  \gamma^0 = \begin{pmatrix} 1 & 0 \\[2pt] 0 & -1 \end{pmatrix}~, \quad 
  \gamma^i  = \begin{pmatrix} 0 & \sigma_i \\[2pt] -\sigma_i & 0 \end{pmatrix}~, \quad
  \gamma^5 = \begin{pmatrix} 0\ &\ 1 \\[2pt] 1\ &\ 0 \end{pmatrix}~.
\end{equation}

%%%%%%%%%%%%%%%%%%%
%%%%%%%%%%%%%%%%%%%
\subsection{The model}
We consider the theory of the relaxion coupled to the SM singlet Dirac fermion through the derivative coupling (see Eq.~(\ref{eq:action:rot})),
\begin{equation}\label{app:eq:Sfer}
 \mathcal {S} = \int d^4 x \sqrt{-g} \Big [ \bar{\psi}\left ( i e^\mu_{\ a} \gamma^a D_\mu - m_\psi - \frac{1}{f_\psi} e^\mu_{\ a} \gamma^a\gamma^5\partial_\mu \phi \right ) \psi + \frac{1}{2} (\partial_\mu \phi )^2 - V(\phi) \Big ]~,
\end{equation}
on the FRW metric
\begin{equation}
  ds^2 = dt^2 - a(t)^2 d{\bf x}^2 =  a^2 \left ( d\tau^2 - d {\bf x}^2 \right )~,
\end{equation}
where $a(t)$ is a scale factor of the Universe. We assume that the pseudo-scalar is spatially homogeneous. More explicit form of the action in Eq.~(\ref{app:eq:Sfer}) on the FRW metric is
\begin{equation}
 \mathcal {S} = \int d\tau d^3{\bf x} \  a^3  \Big [\bar{\psi}\left (  i \gamma^\mu  D_\mu - m_\psi a - \frac{1}{f_\psi} \, \gamma^0 \gamma^5\, \partial_\tau\phi\right ) \psi + \frac{1}{2a} (\partial_\tau \phi )^2 - a V(\phi) \Big ]~.
\end{equation}
The overall scale factor due to $\sqrt{-g}$ in the Lagrangian for fermions can be removed via rescaling, $\psi \rightarrow a^{-3/2} \psi$. Under this rescaling, the covariant derivative due to the spin connection become partial derivative, and the resulting Lagrangian becomes
\begin{equation}\label{app:eq:Lag:Y}
 \mathcal {L} = \bar{\psi}\left ( i \gamma^\mu \partial_\mu  - m_\psi a - \frac{1}{f_\psi} \, \gamma^0\gamma^5\, \partial_\tau{\phi} \right ) \psi + \frac{1}{2} a^2 \left (\partial_\tau{\phi} \right )^2- a^4 V(\phi)~.
\end{equation}
The Lagrangian in Eq.~(\ref{app:eq:Lag:Y}) is problematic for the estimation of the fermion occupation number since the fermionic part in the corresponding Hamiltonian is not entirely captured in the quadratic form in $\psi$:
\begin{equation}
\begin{split}
 {\mathcal H} &= \Pi_\psi\, \partial_\tau\psi + \Pi_\phi\, \partial_\tau\phi - {\mathcal L}~,\\[3pt]
 &= \bar{\psi} \left ( -i \gamma^i \partial_i + m_\psi a + \frac{1}{f_\psi} \gamma^0 \gamma^5 \, \partial_\tau\phi \right ) \psi - \frac{1}{2a^2}\frac{\left (\bar\psi \gamma^0 \gamma^5 \psi \right )^2}{f^2_\psi}  + \frac{1}{2 a^2} \Pi^2_\phi + a^4 V(\phi )~,
\end{split}
\end{equation}
where the quadratic term in $\psi$ is what is taken as the free Hamiltonian in literature and also the part diagonalized by the eigenstates from the equation of motion. 

While the subtlety is linked to the derivative coupling, the derivative coupling can be rotated away by the field redefinition, $\psi \rightarrow e^{-i \frac{\phi}{f_\psi} \gamma^5} \psi$~\cite{Adshead:2018oaa}. In the new basis, the Lagrangian reads 
\begin{equation}\label{app:eq:Lag:Psi}
 \mathcal {L} = \bar{\psi}\left ( i \gamma^\mu \partial_\mu - m_R + i m_I \gamma^5 \right ) \psi + \frac{1}{2} a^2 \left (\partial_\tau {\phi} \right )^2- a^4 V(\phi)~.
\end{equation}
where $m_R = m_\psi\, a \cos \left( \frac{2\phi}{f_\psi} \right)$ and $m_I = m_\psi\, a \sin \left(\frac{2\phi}{f_\psi} \right )$. The corresponding Hamiltonian is given by,  
\begin{equation}
  {\mathcal H} =  \bar{\psi} \left ( - i\gamma^i \partial_i + m_R  - i m_I \gamma^5  \right ) \psi + \frac{1}{2} a^2 \left (\partial_\tau{\phi} \right )^2 + a^4 V(\phi)~.
\end{equation}
Since the Hamiltonian takes a quadratic form in $\psi$, one can unambiguously estimate the fermion occupation number.

%%%%%%%%%%%%%%%%%%%
%%%%%%%%%%%%%%%%%%%
\subsection{Fermion production}
The first step to estimate the fermion occupation number is plugging the expression for the quantum field for $\psi$,
\begin{equation}\label{app:eq:psi:quantum}
\psi = \int \frac{d^3k}{(2\pi)^{3/2}} e^{i{\bf{k}}\cdot {\bf{x}}} \sum_{r=\pm} \left [ U_r( {\bf{k}}, \tau) a_r ({\bf{k}}) + V_r(-{\bf{k}}, \tau) b^\dagger_r (-{\bf{k}})  \right ]~,
\end{equation}
in the Hamiltonian while keeping the relaxion as a classical source. The spinor function in the helicity basis can be parametrized as
\begin{equation}
 U_r ({\bf k}, \tau) = \frac{1}{\sqrt{2}} \begin{pmatrix} u_r \chi_r \\[2pt] r v_r \chi_r \end{pmatrix}~,
 \quad
  V_r ({\bf k}, \tau) = C {\bar U}^T_r \quad \text{with}\quad C = \begin{pmatrix} 0 & i\sigma_2 \\ i\sigma_2 & 0 \end{pmatrix} ~,
\end{equation}
where two-component spinor $\chi_r$ with the helicity $r$ satisfies the eigenvalue equation, $\vec{\sigma}\cdot {\bf k}\, \chi_r = rk\, \chi_r$, and $u_r$ and $v_r$ represent the relative amplitudes between two different chirality states.
The Hamiltonian in terms of the creation and annihilation operators is 
\begin{equation}\label{app:eq:Eenergy:operators}
  {\mathcal{H}} = \sum_{r=\pm}\int d^3k \left ( a^\dagger_r ({\bf k}),\, b_r (-{\bf k})  \right ) \begin{pmatrix} A_r & B^*_r \\[2pt] B_r & - A_r\end{pmatrix} \begin{pmatrix} a_r ({\bf k}) \\[2pt]  b^\dagger_r (-{\bf k})  \end{pmatrix}~,
\end{equation}
where the expressions for $A_r, \ B_r$ in terms of $u_r$ and $v_r$ can be found in~\cite{Adshead:2018oaa},
\begin{equation}
\begin{split}
 A_r &= \frac{1}{2} \Big [ m_R (|u_r|^2 - |v_r|^2) + k (u^*_r v_r + v_r u^*_r) - i r m_I (u^*_r v_r - v^*_r u_r ) \Big ]~,\\[2pt]
 B_r &= \frac{r e^{i r \varphi_{\bf k}}}{2} \Big [ 2 m_R u_r v_r - k (u^2_r - v^2_r) - i r m_I (u^2_r + v^2_r ) \Big ]~,\quad
 e^{i r \varphi_{\bf k}} \equiv \frac{k_1 + i k_2}{\sqrt{k_1^2 + k_2^2}}~.
\end{split}
\end{equation}
The Hamiltonian in Eq.~(\ref{app:eq:Eenergy:operators}) can be diagonalized by the unitary transformation,
\begin{equation}
 \begin{pmatrix} a_r ({\bf k}) \\[2pt]  b^\dagger_r (-{\bf k})  \end{pmatrix} \rightarrow \begin{pmatrix} \alpha^*_r & \beta^*_r \\[2pt] -\beta_r & \alpha_r \end{pmatrix} \begin{pmatrix} a_r ({\bf k}) \\[2pt]  b^\dagger_r (-{\bf k})  \end{pmatrix}~,
\end{equation}
where the mixing angles $\alpha_r$ and $\beta_r$ are called Bogoliubov coefficients which are functions in terms of $u_r$ and $v_r$.
The Hamiltonian in Eq.~(\ref{app:eq:Eenergy:operators}) has two energy eigenvalues,
\begin{equation}
   \pm \omega = \pm \sqrt{k^2 + m^2_R + m^2_I}~.
\end{equation}
The occupation number is given by
\begin{equation}\label{app:eq:nrk:psi}
  n_r(\tau) = |\beta_r|^2 = \frac{1}{2} - \frac{m_R}{\omega}  (|u_r|^2 - |v_r|^2) - \frac{k}{2\omega} Re(u^*_r v_r) - \frac{r m_I}{2\omega} Im (u^*_r v_r)~.
\end{equation}
More details can be found in~\cite{Adshead:2018oaa}. One notes that the fermion occupation number as a function of the cosmic time $t$ is the same as the one in terms of $\tau$. 

Alternatively, the occupation number can be derived purely based on the group theoretic property as recently shown in~\cite{Min:2018rxw}. In the group theoretic approach in~\cite{Min:2018rxw}, the authors showed that the expression in Eq.~(\ref{app:eq:nrk:psi}) collapses into the $SO(3)$~\footnote{$SO(3) \sim SU(2)$ is the subgroup of the symmetry group that corresponds to the freedom in the representation of gamma matrices in Clifford algebra~\cite{Min:2018rxw}.}
invariant form,
\begin{equation}\label{app:eq:nrk:psi:su2}
 n_r(\tau) = \frac{1}{2} \left ( 1 - \frac{{\bf q}\cdot \vec\zeta_r}{|{\bf q}|} \right)~,
\end{equation}
where two $SO(3)$ vectors $\vec{\zeta}_r$ and ${\bf q}$ are given by
\begin{equation}
\begin{split}
 \vec{\zeta}_r &= \frac{1}{2} (u_r^*,\, r\, v^*_r )\, \vec\sigma \begin{pmatrix} u_r \\ r v_r \end{pmatrix} \\
 &= r\, Re(u_r^*v_r)\, \hat{x}_1 + r\, Im (u_r^* v_r)\, \hat{x}_2 + \frac{1}{2} \left ( |u_r|^2 - |v_r|^2 \right )\, \hat{x}_3 \quad \text{with} \quad |\vec\zeta_r| = 1~,
 \\[5pt]
 {\bf q} &= rk\, \hat{x}_1 + m_I\, \hat{x}_2 + m_R\, \hat{x}_3 \quad \text{with}\quad |{\bf q}| = \omega~,
\end{split}
\end{equation}
and they are subject to the equation of motion,
\begin{equation}\label{app:eq:precession}
  \frac{1}{2} \partial_\tau \vec\zeta_r = {\bf q} \times \vec\zeta_r~,
\end{equation}
which looks similar to, for instance, that of the classical precession motion. The equation of motion in Eq.~(\ref{app:eq:precession}) reproduces the same result as the one from the traditional approach.

%%%%%%%%%%%%%%%%%%%
%%%%%%%%%%%%%%%%%%%
\subsection{Solution of equation of motion}
\label{app:sec:eom}

While the fermion production is unambiguously defined with the Hamiltonian from the Lagrangian in Eq.~(\ref{app:eq:Lag:Psi}), the analytic solution is more clearly obtained in the basis with the derivative coupling in Eq.~(\ref{app:eq:Lag:Y}). The equation of motion from the Lagrangian in Eq.~(\ref{app:eq:Lag:Y}) is 
\begin{equation}\label{app:eq:eom:Y}
 \left ( i \gamma^0 \partial_\tau +  i \gamma^i \partial_i - m_\psi a - \frac{a}{f_\psi}  \dot{\phi} \,  \gamma^0\gamma^5 \right ) \psi =0~,
\end{equation}
where dot denotes the differentiation with respect to the cosmic time, for instance, $\dot\phi = \partial_t \phi$.
Using the relation $\vec{\sigma}\cdot {\bf k}\chi_r = rk \chi_r$, the equation of motion in Eq.~(\ref{app:eq:eom:Y}) reduces to those in terms of $\tilde{u}_r$ and $\tilde{v}_r$, and they are
\begin{equation}
\begin{split}
 \begin{pmatrix} i \partial_\tau - m_\psi a \quad & - r\, k - a \displaystyle\frac{\dot\phi}{f_\psi} \\
  r\, k + a \displaystyle\frac{\dot\phi}{f_\psi} & - i \partial_\tau - m_\psi a \end{pmatrix} 
  \begin{pmatrix} \tilde{u}_r \\[5pt]  r \tilde{v}_r \end{pmatrix} =0~, 
\end{split}
\end{equation}
To convert the equation of motion into more familar form where analytic solutions are manifest, we introduce a new set of variables,
\begin{equation}
 x = - k \tau ~, \quad \mu = \frac{m_\psi}{H}~, \quad \xi = \frac{\dot\phi}{2 f_\psi H}~.
\end{equation}
Then, the equation of motion in terms of a new set of variables becomes
\begin{equation}
\begin{split}
 &\left (i \partial_x + \frac{\mu}{x} \right ) \tilde{u}_r + \left ( 1 + \frac{2\xi}{x} r \right ) \tilde{v}_r = 0~,\\[2pt]
 &\left ( i \partial_x - \frac{\mu}{x} \right ) \tilde{v}_r + \left ( 1 + \frac{2\xi}{x} r \right ) \tilde{u}_r = 0~,
\end{split}
\end{equation}
where $\tilde{u}_r$ and $\tilde{v}_r$ were used to refer to the basis with the derivative coupling while keeping $u_r$ and $v_r$ for the basis without the derivative coupling. We take two linear combinations, $\tilde{s}_r = (\tilde{u}_r + \tilde{v}_r)/\sqrt{2}$ and $\tilde{d}_r = (\tilde{u}_r - \tilde{v}_r)/\sqrt{2}$
, and iterate two coupled first-order equations to get two decoupled second-order differential equations,
\begin{equation}
\begin{split}
  \partial^2_x \tilde{s}_r + \frac{1}{x}\partial_x \tilde{s}_r + \left [ \left ( 1 + \frac{2\xi}{x} r \right )^2 + \frac{\mu^2}{x^2} - \frac{i}{x} \right ] \tilde{s}_r &= 0~, \\[3pt]
  \partial^2_x \tilde{d}_r + \frac{1}{x}\partial_x \tilde{d}_r + \left [ \left ( 1 + \frac{2\xi}{x} r \right )^2 + \frac{\mu^2}{x^2} + \frac{i}{x} \right ] \tilde{d}_r &= 0~.
\end{split}
\end{equation}
After making a rescaling of $\tilde{s}_r = x^{-1/2} s_r$ (and $\tilde{d}_r = x^{-1/2} d_r$) and changing the variable, $x = -z/(2 i)$, the  differential equations take the form of the Whittaker equation.
\begin{equation}\label{eq:whittaker:eq}
\begin{split}
  \partial^2_z s_r + \left [-\frac{1}{4} +\frac{1}{z} \left ( \frac{1}{2}  + i 2 \xi r \right ) + \frac{1}{x^2} \left ( \frac{1}{4} + \mu^2 + 4\xi^2 \right ) \right ] s_r &= 0~, \\[3pt]
  \partial^2_z d_r + \left [-\frac{1}{4} +\frac{1}{z} \left ( -\frac{1}{2}  + i 2 \xi r \right ) + \frac{1}{x^2} \left ( \frac{1}{4} + \mu^2 + 4\xi^2 \right ) \right ] d_r &= 0~.
\end{split}
\end{equation}
We choose the boundary conditions such that the fermion occupation number in Eq.~(\ref{app:eq:nrk:psi}) vanishes in the limit $x\rightarrow \infty$ (or equivalently $\tau \rightarrow - \infty$ or $t \rightarrow - \infty$ in terms of the conformal time or the cosmic time). One notes that fermion occupation numbers in both basis (with and without the derivative couplings) become identical in the limit $x\rightarrow \infty$ (see section 5 of~\cite{Min:2018rxw} for the detail). Zero occupation number in the basis with the derivative coupling guarantees the same boundary condition in the basis without the derivative coupling. Therefore, we can safely take the solutions obtained in the basis with the derivative coupling, namely
\begin{equation}\label{app:eq:sol:srdr}
\begin{split}
s_r(x) = e^{-\pi r\xi} W_{\frac{1}{2}+2ir\xi, i\sqrt{\mu^2 +4\xi^2}} (-2ix)~, \\
d_r(x) = -i\mu e^{-\pi r\xi} W_{-\frac{1}{2}+2ir\xi, i\sqrt{\mu^2 +4\xi^2}} (-2ix)~,
\end{split}
\end{equation}
and use them in the new basis.

On the other hand, the scanning process in the relaxation mechanism occurs during the finite time, roughly $\Delta t \sim N_e/H$. While it implies that the initial condition for the zero occupation number should be imposed in principle at a finite time $\tau_0$ ($\neq - \infty$), the approximate solutions in Eq.~(\ref{app:eq:sol:srdr}) must be sufficient since the above time interval in the cosmic time corresponds to the exponentially separated time interval in the conformal time,
\begin{equation}\label{eq:app:tautau0}
\frac{\tau}{\tau_0} \sim e^{-H\cdot \Delta t} \sim e^{-N_e} \ll 1 \quad \text{for}\quad N_e \sim \mathcal{O}(10^{1\sim 3})~.
\end{equation}
Although we have checked numerically that the result assuming the finite $\tau_0$ is similar to the asymptotic case with $\tau_0 \rightarrow -\infty$, we provide a semi-analytic proof for the validity of our approximation in Eq.~(\ref{app:eq:sol:srdr}).

\begin{figure}[tph]
\begin{center}
\includegraphics[width=0.48\textwidth]{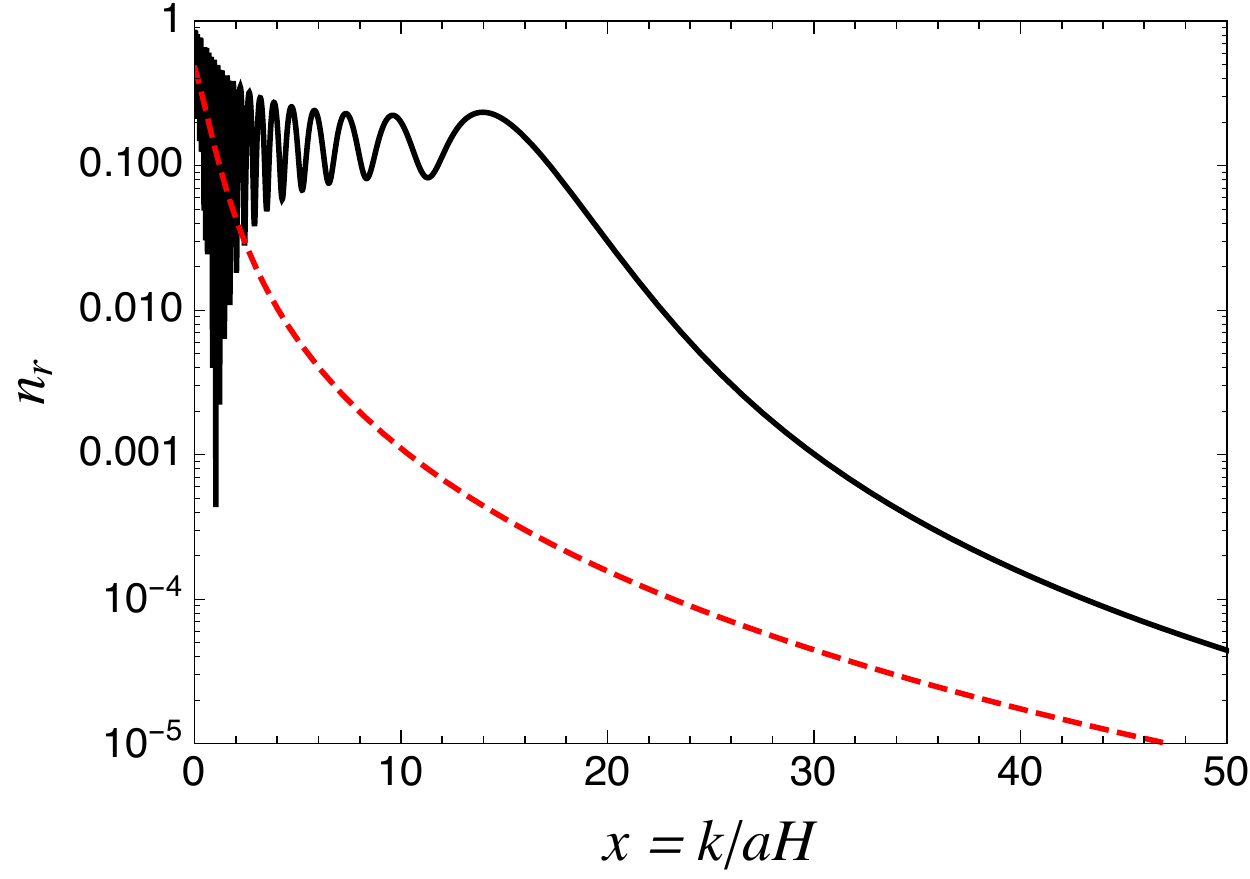}\quad
\includegraphics[width=0.48\textwidth]{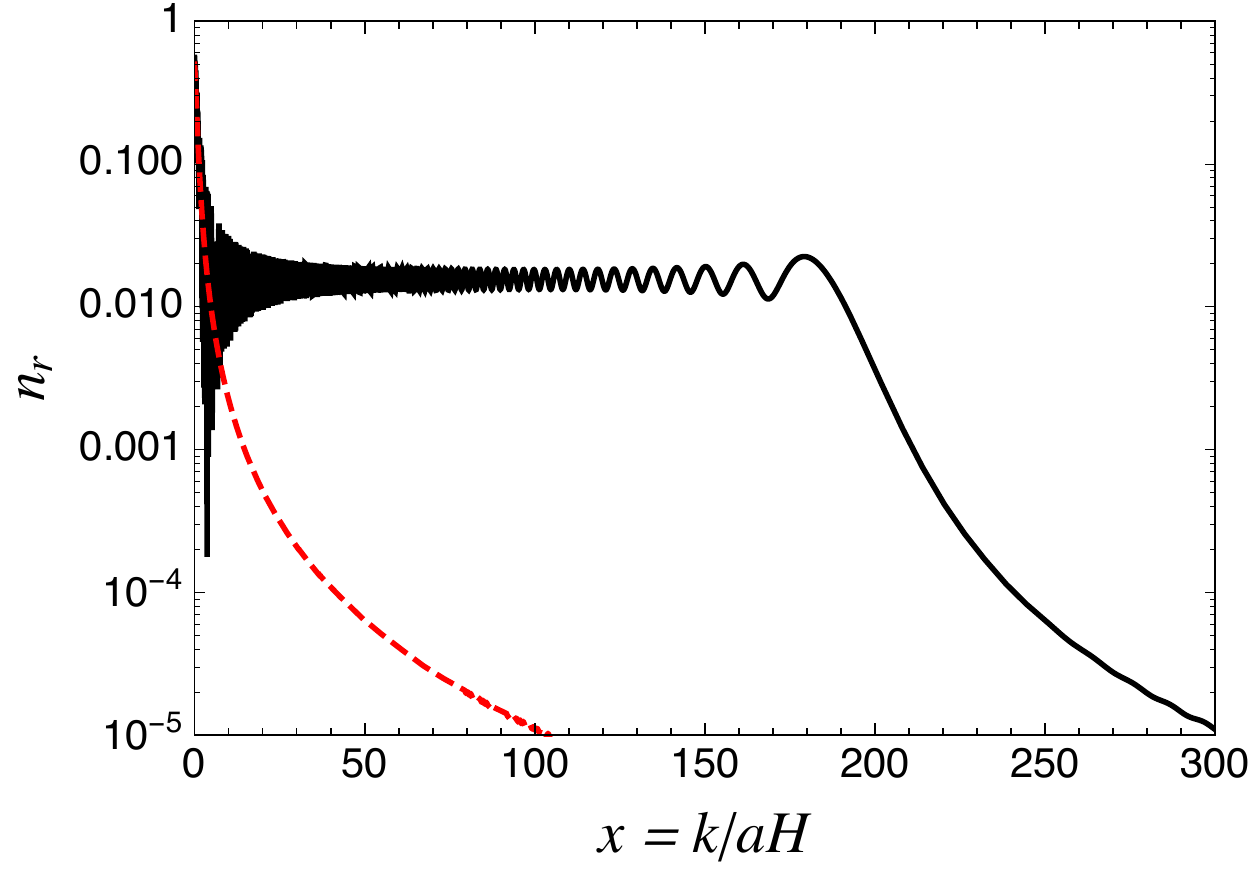}
\\[5pt]
\includegraphics[width=0.48\textwidth]{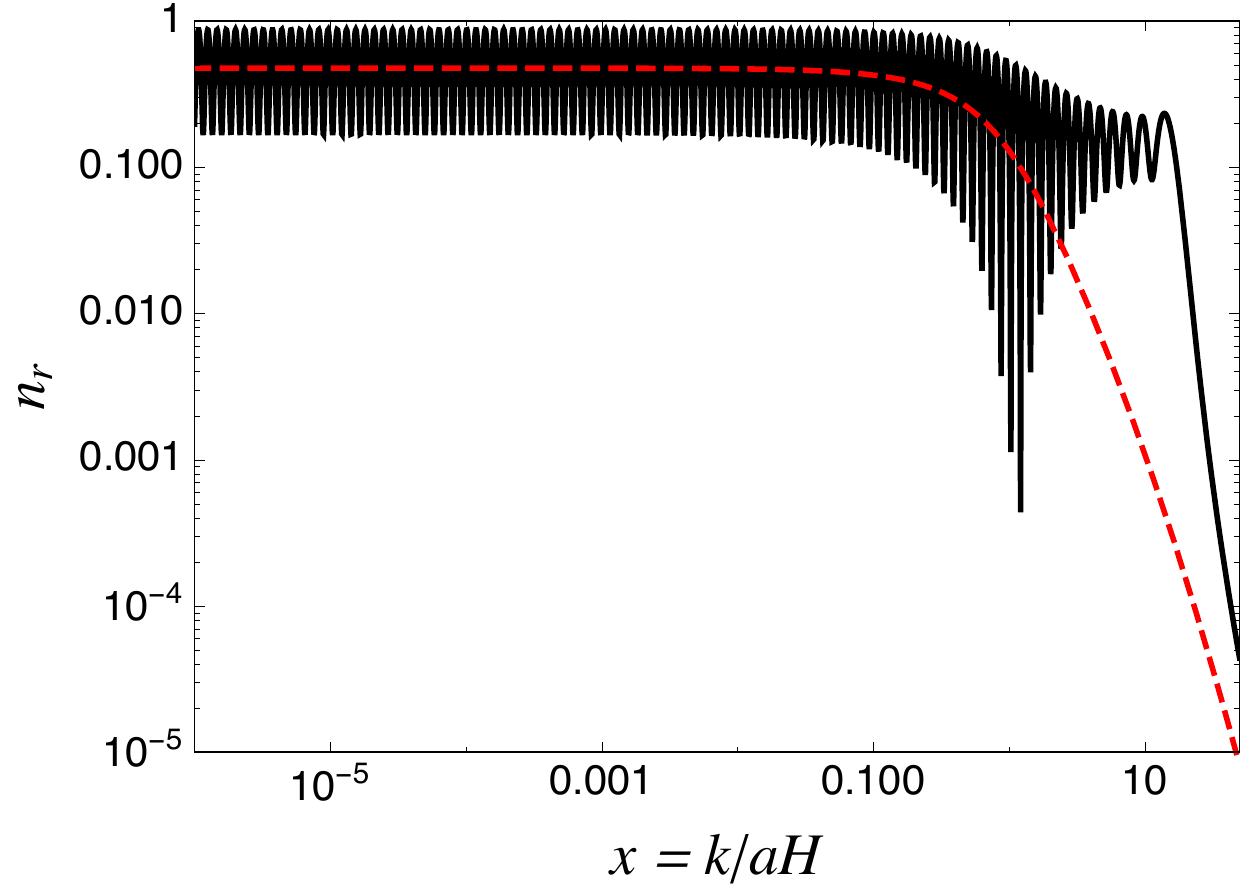}\quad
\includegraphics[width=0.48\textwidth]{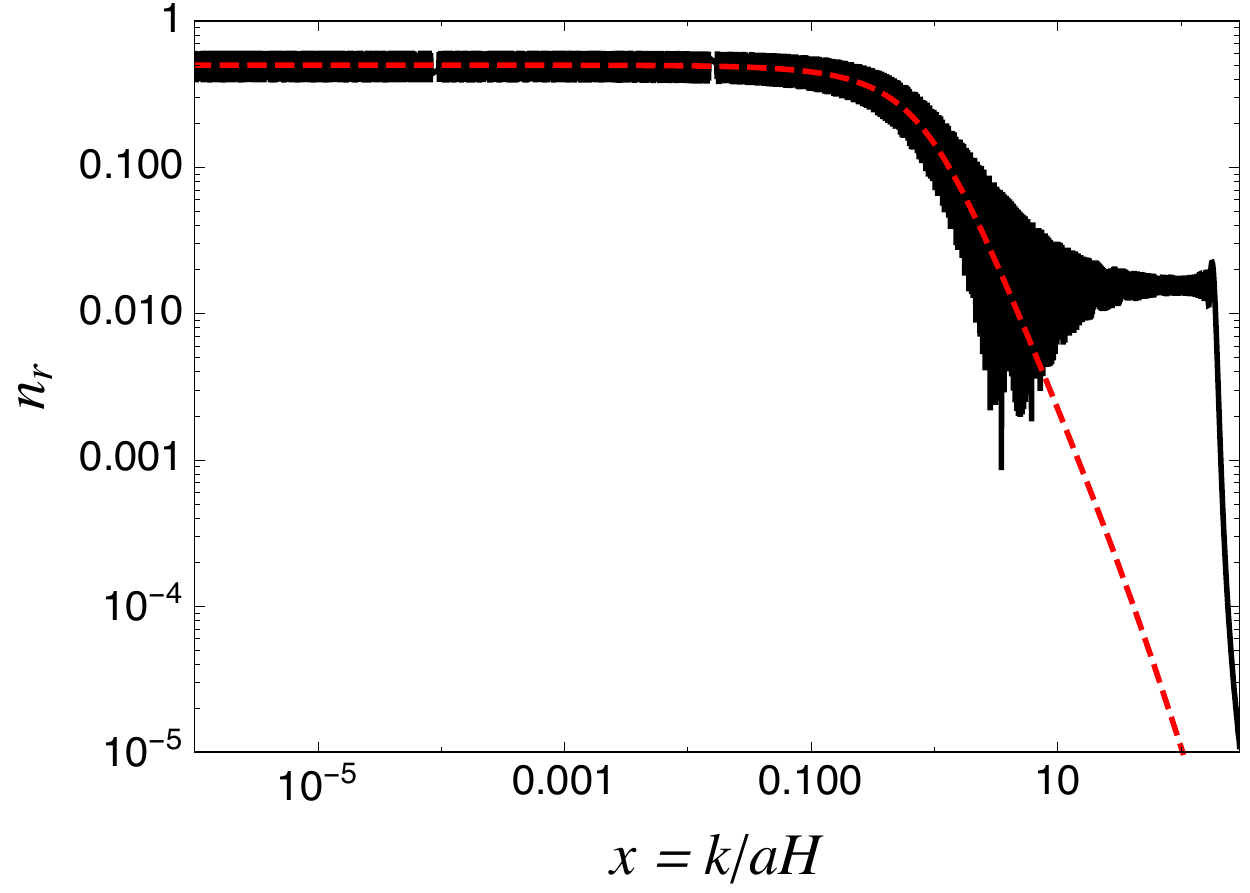}
\caption{\small The occupation number (see Eq.~(\ref{app:eq:nrk:psi}) or (\ref{app:eq:nrk:psi:su2})) for the helicity $r=-1$ (solid black) and $r=1$ (dashed red) as a function of $x=k/aH$. The plots correspond to $\xi = 10$ (left) and $\xi = 100$ (right) with $\mu=1$ for the purpose of illustraion. The same plots are displayed in two different styles: typical scale (top) and logarithmic scale (bottom) of the horizontal axes.}
\label{fig:occupation:typ:log}
\end{center}
\end{figure}

Since the velocity $\dot\phi$ is positive, the occupation number is dominated by fermions with the helicity $r=-1$ since the fermion  production can actively occur whenever 
\begin{equation}\label{app:eq:effk}
 r\, k + a\, \frac{\dot\phi}{f_\psi} = 0~,
\end{equation} 
and it is easily satisfied for the helicity $r=-1$ around $k\sim -2\xi/\tau$. Away from $k\sim -2\xi/\tau$, the WKB approximation is valid, and the occupation number is roughly constant. Even in the absence of the interaction of the relaxion to fermions, the occupation number is non-vanishing since a free massive fermion system in the expanding Universe can cause the fermion production. The latter contribution is not captured by the relation in Eq.~(\ref{app:eq:effk}) as it is a gravitational effect, and we find that it shuts off around $k\sim -\mu/\tau$. Consequently, the occupation number for the helicity $r=-1$, assuming an initial condition at $\tau_0 \rightarrow -\infty$, can be written approximately as
\begin{eqnarray}\label{eq:nr:approx}
n_r(k) & \approx &
\left\{ 
\begin{array} {lll}
\displaystyle\frac{1}{2} &\hspace{0.3cm} {\rm for} & \hspace{0.3cm} k < \displaystyle\frac{\mu}{-\tau} \\[13pt]
\displaystyle\frac{\mu^2}{\xi} &\hspace{0.3cm} {\rm for} &\hspace{0.3cm} \displaystyle\frac{\mu}{-\tau} < k < \displaystyle\frac{2\xi}{-\tau} = 2\xi H a\\[13pt]
0 &\hspace{0.3cm} {\rm for} &\hspace{0.3cm} k > \displaystyle\frac{2\xi}{-\tau} ~,
\end{array}
\right.
\end{eqnarray}
where the momentum $k$ is the one in the frame with the conformal time and $\mu^2/\xi$ in the region of interest is typically much smaller than 1/2.
The numerical validation of the approximated occupation number in Eq.~(\ref{eq:nr:approx}) is shown in Fig.~\ref{fig:occupation:typ:log} for two choices of $\xi$ values, $\xi = 10,\, 100$, for the purpose of illustration (see also Fig.~1 of~\cite{Adshead:2018oaa} or Fig.~4 of~\cite{Min:2018rxw} for related discussions). While the numerical simulation with a larger $\xi$ in our benchmark points is technically challenging due to the highly oscillatory behavior, we suspect the generic feature remains the same. 

%%%%%%%%%%%%%%%%%%%%%%
%%%%%%%%%%%%%%%%%%%%%%

When the zero occupation number is imposed at a finite time $\tau_0$ ($\neq -\infty$), the solution of the Whittaker equation in Eq.~(\ref{eq:whittaker:eq}) is given in terms of the Whittaker functions of both kinds (instead of the approximation in Eq.~(\ref{app:eq:sol:srdr})),
\begin{equation}\label{eq:sd:fintetime}
 s, d (x) = A\, W^{(1)}(x) + B\, W^{(2)}(x)~,
\end{equation} 
where $W^{(1)}$ ($W^{(2)}$) is the Whittaker function of the first (second) kind. The coefficients $A,\, B$ in Eq.~(\ref{eq:sd:fintetime}) are found to be functions of three dimensionless parameters,
\begin{equation}
  x_0 = - k \tau_0~, \quad 2 \xi = \frac{\dot\phi}{f_\psi H}~,\quad \mu = \frac{m_\psi}{H}~.
\end{equation}
It can be shown that $A,\, B$ converge into their asymptotics at $x_0 \rightarrow \infty$ (as those in Eq.~(\ref{app:eq:sol:srdr})) as long as $x_0 = - k\tau_0 \gg 2 \xi$. Since $\tau/\tau_0 \ll 1$ and the parametrization in Eq.~(\ref{eq:nr:approx}), the contribution to the fermion production from the momentum interval $k= (-2\xi/\tau_0, \, -2\xi/\tau)$ will agree with our approximation with Eq.~(\ref{app:eq:sol:srdr}). It is the contribution from the interval $k= (0, \, -2\xi/\tau_0)$ that might potentially invalidate our approximation. 

Defining an intermediate momentum scale,
\begin{equation}
k_{int} = {\rm max} (-\mu/\tau,\, -2\xi/\tau_0)~,
\end{equation}
and taking $n_r (k < k_{\rm int})=1$ as a conservative choice, our approximation with the solutions in Eq.~(\ref{app:eq:sol:srdr}) will be valid as long as the following inequality for the number densities integrated over each momentum interval is satisfied,
\begin{equation}\label{app:eq:inequality}
\int_0^{k_{int}} d^3k\, n_r (k) \approx k_{int}^3~ \ll~
\int_{k_{int}}^{-2\xi/\tau} d^3k\, n_r (k) \approx \frac{\mu^2}{\xi} \left ( \frac{2\xi}{-\tau} \right )^3~,
\end{equation} 
where the contributions are dominated by the case with $r=-1$ and we dropped the factor from the integration over the solid angle. Demanding inequality in Eq.~(\ref{app:eq:inequality}) gives rise to the constraint, 
\begin{equation}\label{eq:approx;validation}
\begin{array} {lll}
\mu \ll \xi^2 &\hspace{0.3cm} {\rm when} & \hspace{0.3cm} k_{int} = \displaystyle\frac{\mu}{-\tau}  \\[13pt]
\displaystyle\frac{\xi}{\mu^2} \ll \left ( \frac{\tau_0}{\tau} \right )^3 = e^{3N_e} &\hspace{0.3cm} {\rm when} &\hspace{0.3cm} k_{int} = \displaystyle\frac{2\xi}{-\tau_0}~.
\end{array}
\end{equation}
The constraints in Eq.~(\ref{eq:approx;validation}) are satisfied in our benchmark points as long as the number of e-folding is bigger than $\sim \mathcal{O}(10)$, namely $N_e \gtrsim \mathcal{O}(10)$. Therefore, we conclude that our approximation using the solutions in Eq.~(\ref{app:eq:sol:srdr}) is justified.

%%%%%%%%%%%%%%%%%%%
%%%%%%%%%%%%%%%%%%%
\subsection{Backreaction}

The equation of motion for the relaxion in presence of the coupling to the fermion $\psi$ is obtained in the basis without the derivative coupling, and it is given by
\begin{equation}
\ddot{\phi}+3 H \dot{\phi}+\frac{\partial}{\partial \phi} V(\phi ) = \frac{2m_\psi}{f_\psi a^3} \bar{\psi}\left[ \sin \left( \frac{2\phi}{f_\psi} \right) + i \gamma^5 \cos \left( \frac{2\phi}{f_\psi} \right) \right] \psi~,
\end{equation}
where the term in the right-hand side is the backreaction, what we call $\mathcal B$, due to the fermion production. Using the expression for the quantum field in Eq.~(\ref{app:eq:psi:quantum}), it is given by in terms of $u_r$ and $v_r$ (also in terms of $\tilde{u}_r$ and $\tilde{v}_r$, quantities in the basis with the derivative coupling),
\begin{equation}
\begin{split}\label{app:eq:br:kspace}
{\mathcal B} &= \frac{m_\psi}{ f_\psi a^3} \sum_r \int \frac{d^3 k}{(2\pi)^3} \Big [ \sin \left ( \frac{2\phi}{f_\psi} \right ) \left ( |v_r|^2 - |u_r|^2 \right ) - i r \cos \left ( \frac{2\phi}{f_\psi} \right ) \left ( u^*_r v_r - u_r v^*_r \right ) \Big ]~,\\[2.5pt]
&=\frac{m_\psi}{ f_\psi a^3} \sum_r \int \frac{d^3 k}{(2\pi)^3} \Big [ -i r (\tilde{u}^*_r \tilde{v}_r - \tilde{u}_r \tilde{v}^*_r) \Big ]~.
\end{split}
\end{equation}
The above expression differs from that in~\cite{Adshead:2018oaa} by the overall factor of 2. We suspect that it is due to the missed normalization factor $1/\sqrt{2}$ in $V_r({\bf k},\, t)$.
Using the relations, $\tilde{u}_r = (s_r + d_r )/\sqrt{2x}$ and $\tilde{v}_r = (s_r - d_r )/\sqrt{2x}$ in Section~\ref{app:sec:eom}, the backreaction can be expressed in terms of $s_r$ and $d_r$ whose analytic solutions were the Whittaker functions,
\begin{equation}
\begin{split}\label{app:eq:br:xspace}
{\mathcal B} 
&= \frac{m_\psi H^3}{\pi^2 f_\psi} \sum_r r \int^\infty_0 dx \, x\, \Im \left[ d^*_r(x) s_r (x) \right]~.
\end{split}
\end{equation}
The integration is non-trivial, but it can be done analytically using the integral representation of the Whittaker functions known as the Mellin-Barnes representation,
\begin{equation} 
W_{ \mathfrak{a},\, \mathfrak{b}}(z) = \frac{e^{-z/2}}{2\pi i} \int_C dt \ \frac{\Gamma \left( \frac{1}{2}+\mathfrak{b}+t \right) \Gamma \left( \frac{1}{2}-\mathfrak{b}+t \right) \Gamma \left( -\mathfrak{a} -t \right)}{\Gamma \left( \frac{1}{2}+\mathfrak{b}-\mathfrak{a} \right) \Gamma\left( \frac{1}{2}-\mathfrak{b}-\mathfrak{a} \right)} z^{-t},
\end{equation}
where the contour $C$ covers from $-i \infty$ to $i \infty$ with a deformation to separate the poles of $\Gamma \left( \frac{1}{2}+\mathfrak{b}+t \right) \Gamma \left( \frac{1}{2}-\mathfrak{b}+t \right)$ from those of $\Gamma \left( -\mathfrak{a} -t \right)$. We have checked the result in~\cite{Adshead:2018oaa}. While we have a few disagreements with those in~\cite{Adshead:2018oaa} in some detail (possibly due to typos), we have reproduced the same result as~\cite{Adshead:2018oaa} up to a factor of 2 that was mentioned above.
Since all necessary details can be found in~\cite{Adshead:2018oaa}, we do not repeat the computation here. Instead, we present only the final result for the backreaction. It is given by 
\begin{equation}\label{app:eq:backreaction}
\begin{split}
{\mathcal B} =&\ \frac{m_\psi H^3}{\pi^2 f_\psi}\sum_r r \int^\Lambda_0 dx \, x\, \Im \left[ d^*_r(x) s_r (x) \right] 
\\[2.5pt]
=&\  \frac{\mu^2 H^4}{\pi^2 f_\psi} \Big [ -6\xi\left( \ln 2\Lambda +\gamma_E \right) -\frac{3}{2}\sqrt{\mu^2+4\xi^2}
\sinh(4\pi \xi)\text{csch}(2\pi \sqrt{\mu^2+4\xi^2}) +7 \xi
\\[2.5pt]
& +\frac{i}{4}(\mu^2-8\xi^2-6i\xi+1)  
\cdot \Big [ \left(1+\sinh(4\pi \xi) \text{csch}(2\pi \sqrt{\mu^2+4\xi^2}) \right ) H_{-i(\sqrt{\mu^2+4\xi^2}+2\xi)}
\\[2.5pt]
& +\left(1- \sinh(4\pi \xi) \text{csch}(2\pi \sqrt{\mu^2+4\xi^2}) \right)H_{i(\sqrt{\mu^2+4\xi^2}-2\xi)} \Big ] 
\\[2.5pt]
& -\frac{i}{4}(\mu^2-8\xi^2+6i\xi+1) 
\cdot \Big [ \left(1+ \sinh(4\pi \xi) \text{csch}(2\pi \sqrt{\mu^2+4\xi^2}) \right) H_{i(\sqrt{\mu^2+4\xi^2}+2\xi)}
\\[2.5pt]
&  +\left(1- \sinh(4\pi \xi) \text{csch}(2\pi \sqrt{\mu^2+4\xi^2}) \right)H_{-i(\sqrt{\mu^2+4\xi^2}-2\xi)} \Big ] \ \Big ]~.
\end{split}
\end{equation}

When $\mu^2 \ll \xi $ and $1\ll \xi$ (note that this limit is consistent with the parametrization in Eq.~(\ref{eq:nr:approx})), the backreaction in Eq.~(\ref{app:eq:backreaction}) is approximately given by 
\begin{equation}
\label{app:eq:limit1}
\mathcal{B} \approx -\frac{4}{\pi} \frac{H^4 \mu^2}{f_\psi} \xi^2~,
\end{equation}
where the negative sign implies that the backreaction plays a role of drag force in the classical picture. 
Unlike what was discussed in~\cite{Adshead:2018oaa}, we find that the condition $\mu \ll 1$ is not necessary to derive the approximation  in Eq.~(\ref{app:eq:limit1}). Since $\xi \gg 1$, the condition $\mu^2 \ll \xi$ that leads to the approximation in Eq.~(\ref{app:eq:limit1}) is always satisfied for $\mu \ll 1$.
Although the approximation in Eq.~(\ref{app:eq:limit1}) apparently does not depend on the Hubble parameter $H$ as $\mu,\, \xi \propto 1/H$, it should not be considered to be valid in the $H \rightarrow 0$ limit.
Since two expansion parameters in terms of $H$ scale like $1/\xi \propto H$ and $\mu^2/\xi \propto 1/H$, the limit that leads to the approximation in Eq.~(\ref{app:eq:limit1}) is not consistent with the $H \rightarrow 0$ limit.

\begin{figure}[tph]
\begin{center}
\includegraphics[width=0.48\textwidth]{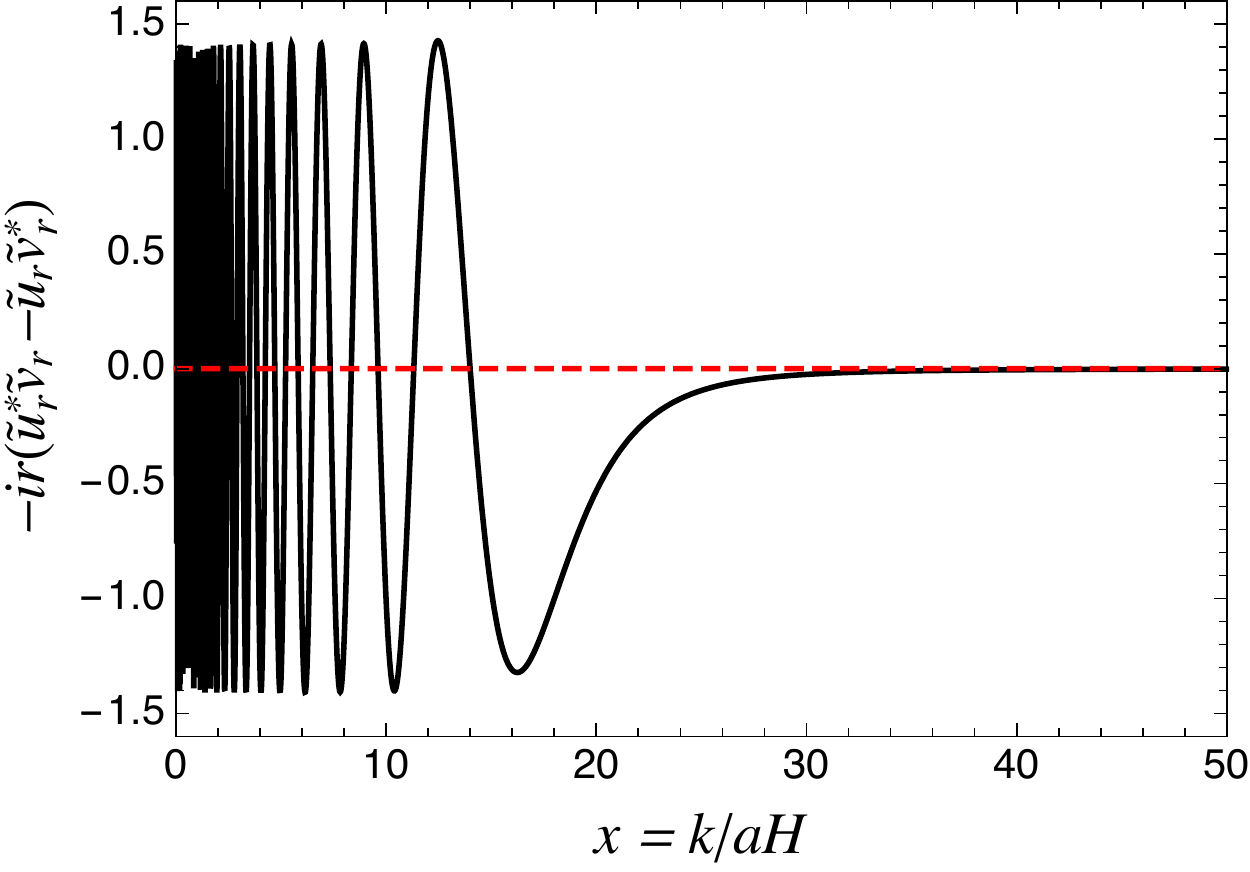}\quad
\includegraphics[width=0.48\textwidth]{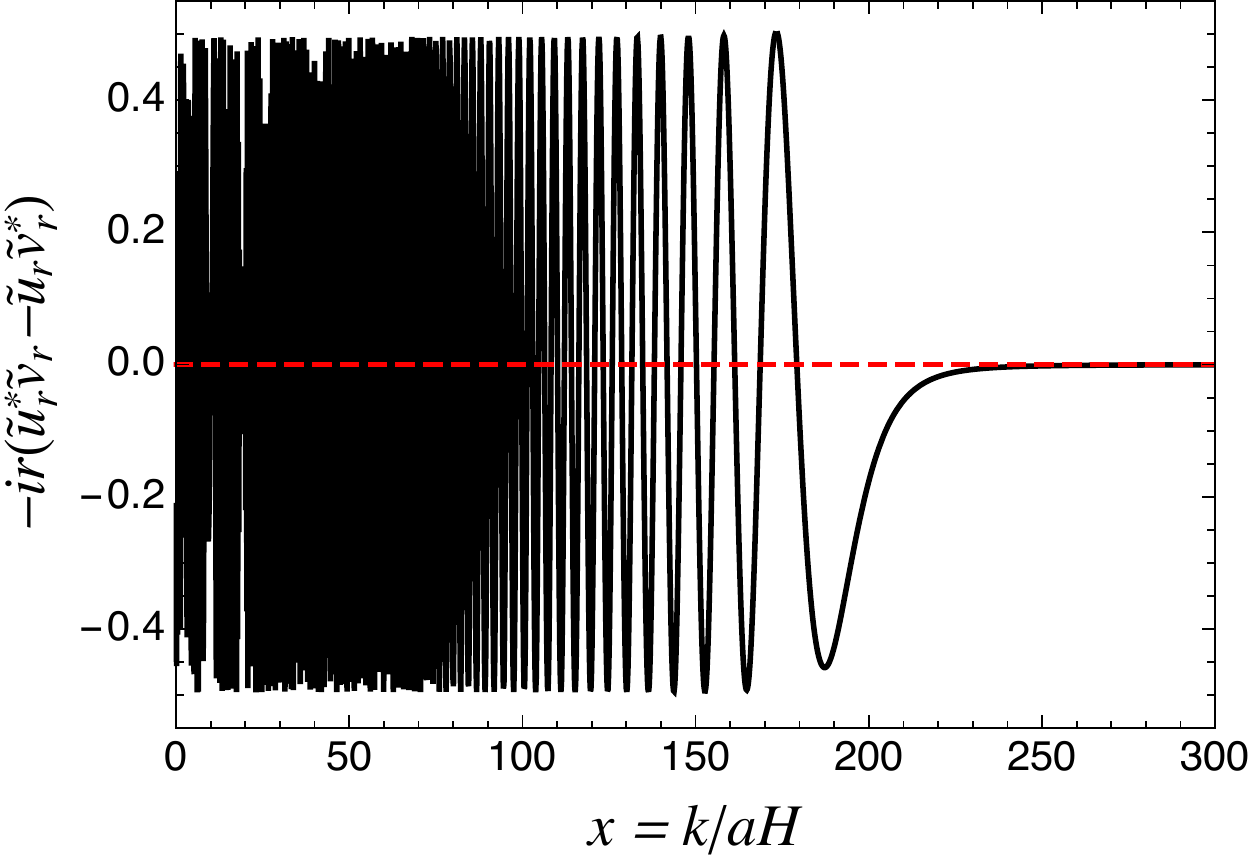}
\\[5pt]
\includegraphics[width=0.48\textwidth]{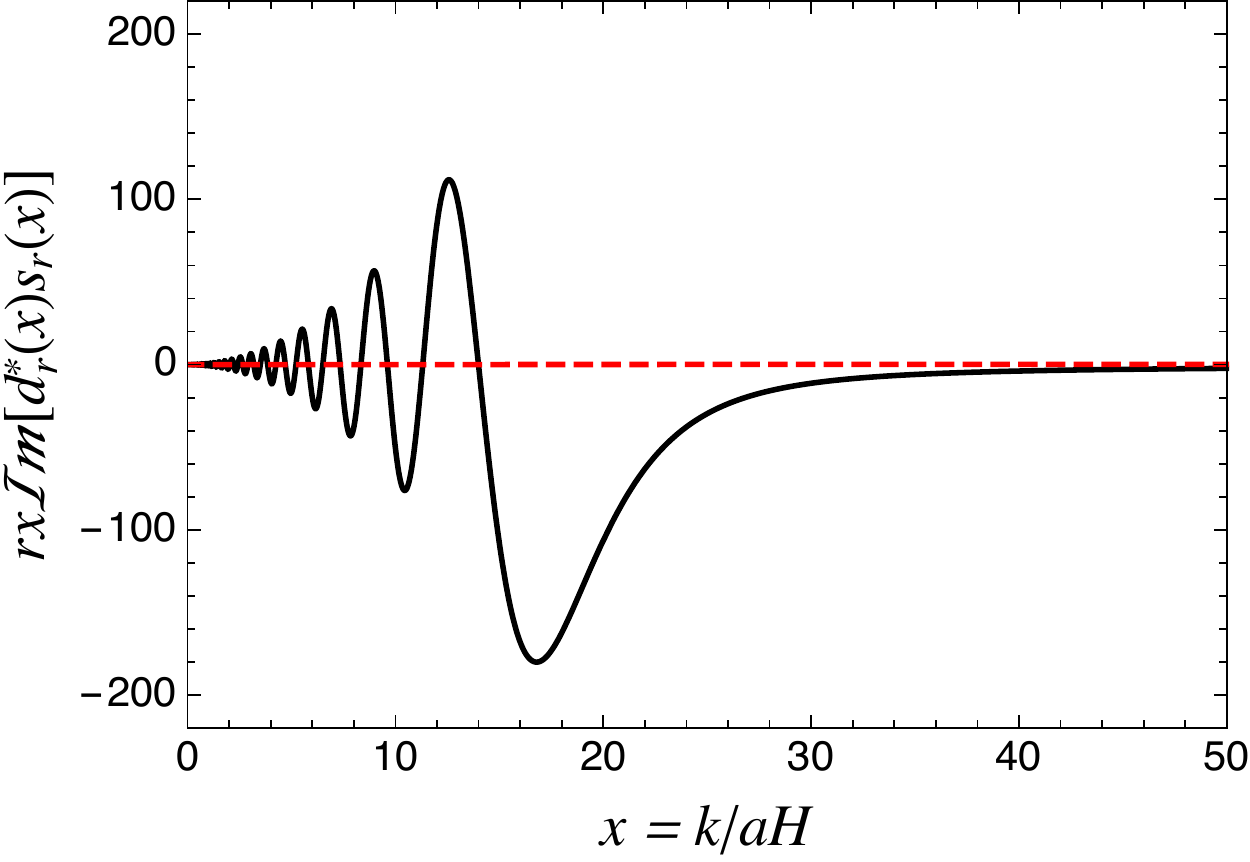}\quad
\includegraphics[width=0.48\textwidth]{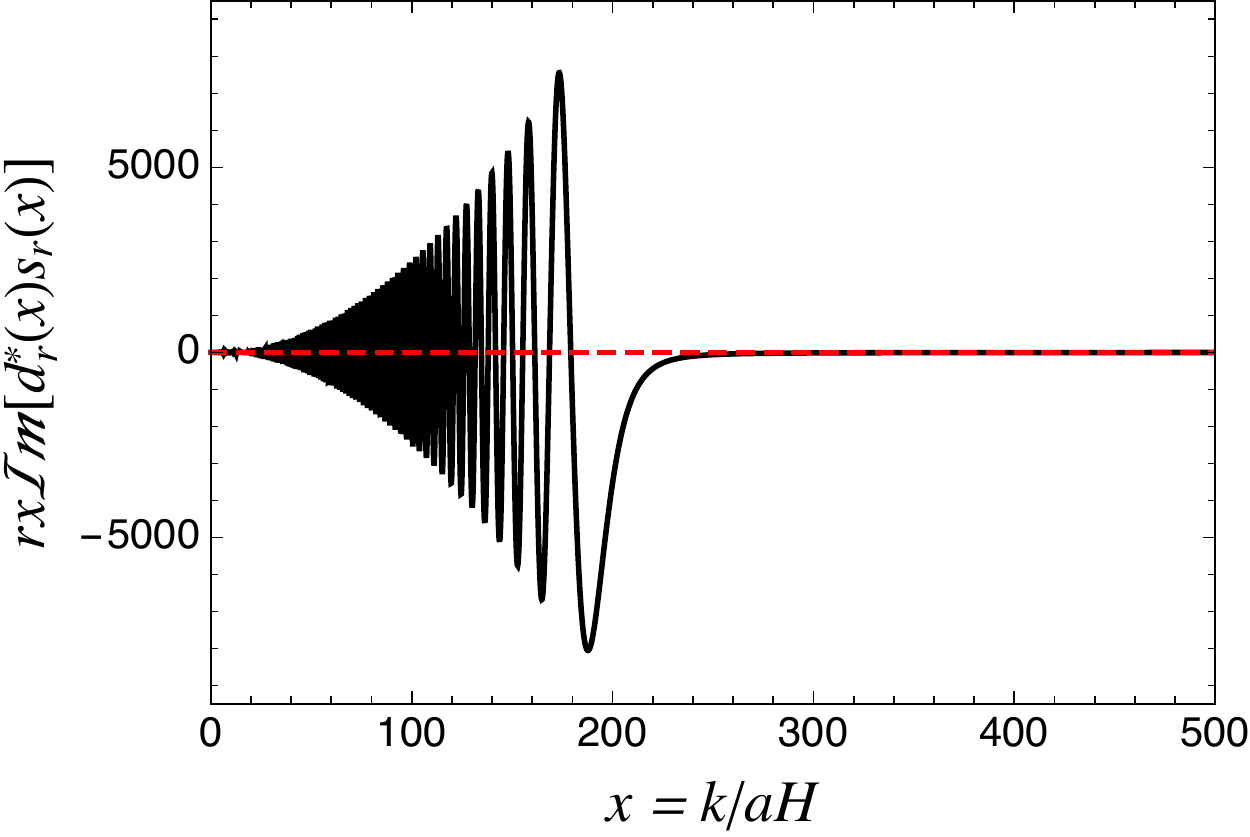}
\caption{\small The integrand of the backreaction given in Eq.~(\ref{app:eq:br:kspace}) (upper panel) and Eq.~(\ref{app:eq:br:xspace}) (lower panel) for the helicity $r=-1$ (solid black) and $r=1$ (dashed red) as a function of $x=k/aH$. The plots correspond to $\xi = 10$ (left) and $\xi = 100$ (right) with $\mu=1$ for the purpose of illustration.}
\label{fig:backreaction}
\end{center}
\end{figure}

For the purpose of illustration, we numerically evaluate the integrand of the backreaction in Eq.~(\ref{app:eq:br:kspace}), namely before the integration over the momentum, for the same set of $\xi$ values as those in Fig.~\ref{fig:occupation:typ:log}, $\xi = 10,\, 100$ (smaller than our benchmark values), and they are shown in the upper panel of Fig.~\ref{fig:backreaction} as a function of $x = k/aH$. While the $x$-dependence in $d^3k/a^3$ is not included in the upper panel of Fig.~\ref{fig:backreaction}, including it amounts to simulate the integrand of Eq.~(\ref{app:eq:br:xspace}). As is evident in the lower panel of Fig.~\ref{fig:backreaction}, it is more pronounced in a larger $x$ region though due to the multiplicative $x$.  Similarly to the occupation number in Fig.~\ref{fig:occupation:typ:log}, the backreaction shuts off around $x \sim 2 \xi$ as is expected.

%%%%%%%%%%%%%%%%%%%%%%
%%%%%%%%%%%%%%%%%%%%%%

%%%%%%%%%%%%%%%%%%%
%%%%%%%%%%%%%%%%%%%
\subsection{Fermion energy density}

In this section, we estimate the fermion energy density using the previously obtained fermion occupation number. 
Unlike the case of the backreaction, we provide the full computation filling the gap in~\cite{Adshead:2018oaa}.
The total energy density summed over the fermion and its anti-particle is given by
\begin{equation}
\begin{split}
\rho_\psi (\tau) & = 2 \sum_r \int d^3k \ \omega\, n_r (\tau)~.
\end{split}
\end{equation}
We can re-express $\rho_\psi (\tau)$ in terms of $s_r$ and $d_r$ that we know their solutions,
\begin{equation}\label{app:eq:rhot:exp2}
\begin{split}
\rho_\psi (\tau) 
& = \frac{8\pi}{\tau^4} \sum_r \int^\Lambda_0  dx\, x^2 \ \left[\frac{\sqrt{x^2+\mu^2}}{2}+\frac{x}{2}-\frac{\mu}{2x} \Re[s^*_r d_r] -\frac{|s_r|^2}{2} \right] ~,
\\[2.5pt]
& = \frac{4\pi}{\tau^4} \sum_r \left[  \int^\Lambda_0 dx \ \left( x^2 \sqrt{x^2+\mu^2} +x^3 \right) -\mu  \int^\Lambda_0 dx \, x\, \Re[s^*_r d_r] -  \int^\Lambda_0 dx \ x^2 | s_r |^2 \right ]~.
\end{split}
\end{equation}
In what follows, we will calculate each contribution in the above expression individually, providing the detail.

The first term of the second line in Eq.~(\ref{app:eq:rhot:exp2}) is trivially evaluated, and it is given by
\begin{equation}
\begin{split}
& \sum_r \int^\Lambda_0 dx \ \left( x^2 \sqrt{x^2+\mu^2} +x^3 \right) \\
& \quad = \frac{1}{4} \Lambda (2 \Lambda^2 +\mu^2) \sqrt{\Lambda^2 + \mu^2}+\frac{1}{4} \mu^4 \ln \frac{\mu}{\Lambda + \sqrt{\Lambda^2 + \mu^2}} +\frac{\Lambda^4}{2}~.
\end{split}
\end{equation}
In the limit $\Lambda \gg \mu$, it is approximated to be
\begin{equation}
\Lambda^4+\frac{1}{2}\mu^2 \Lambda^2+\frac{\mu^4}{16} + \frac{\mu^4}{4} \ln \frac{\mu}{2\Lambda} ~.
\end{equation}
The second term of the second line in Eq.~(\ref{app:eq:rhot:exp2}) is the real part of what we have already calculated in the backreaction (except the helicity $r$ in the helicity sum):
\begin{equation}
\begin{split}
\mu  \sum_r & \int^\Lambda_0 dx\, x\, \Re[s^*_r d_r] \\
= & \ \frac{\mu^2}{2} \Big[ 2\Lambda^2 -2(\mu^2-2 \mathfrak{b}^2+1)(\ln 2\Lambda +\gamma_E) +\mu^2 +4 -6 \mathfrak{b}^2 +6 \mathfrak{a}\mathfrak{b}\cdot \frac{\sinh (2\pi \mathfrak{b})}{\sinh (2\pi \mathfrak{a})} \\
& +\frac{1}{2} \left( 1- \frac{\sinh (2\pi \mathfrak{b})}{\sinh (2\pi \mathfrak{a})} \right) \left[ (\mu^2 -2 \mathfrak{b}^2 +3i \mathfrak{b}+1) H_{-i(\mathfrak{a}-\mathfrak{b})}+(\mu^2 -2 \mathfrak{b}^2 -3i \mathfrak{b}+1) H_{i(\mathfrak{a}-\mathfrak{b})} \right] \\
& +\frac{1}{2} \left( 1+ \frac{\sinh (2\pi \mathfrak{b})}{\sinh (2\pi \mathfrak{a})} \right) \left[ (\mu^2 -2 \mathfrak{b}^2 +3i \mathfrak{b}+1) H_{i(\mathfrak{a}+\mathfrak{b})}+(\mu^2 -2 \mathfrak{b}^2 -3i \mathfrak{b}+1) H_{-i(\mathfrak{a}+\mathfrak{b})} \right] \ \Big] ~,
\end{split}
\end{equation}
where
\begin{equation}
 \mathfrak{a} \equiv \sqrt{\mu^2 + 4 \xi^2}~, \quad \mathfrak{b} \equiv 2 \xi~.
\end{equation}
The third term of the second line in Eq.~(\ref{app:eq:rhot:exp2}) is the volume integration of $|s_r|^2$. Using the integral representation of the Whittaker function, we can the analytic expression for it. 
\begin{equation}\label{app:eq:rho:3rd}
\begin{split}
\sum_r & \int^\Lambda_0 dx \, x^2  |s_r|^2 \\
= \ & \sum_r \int^\Lambda_0 dx \, x^2  e^{-\pi r \mathfrak{b}} \ W_{\frac{1}{2}+ir \mathfrak{b}, i \mathfrak{a}} (-2ix) \ W_{\frac{1}{2}-ir \mathfrak{b}, -i \mathfrak{a}} (2ix) 
\\[2.5pt]
= \ & \sum_r \left(-\frac{e^{-\pi r \mathfrak{b}}}{4 \pi^2} \right) \int_{C'} ds \ \frac{ \Gamma \left( \frac{1}{2} -i \mathfrak{a} +s \right)  \Gamma \left( \frac{1}{2} +i \mathfrak{a} +s \right) \Gamma \left( -\frac{1}{2} +ir \mathfrak{b} -s \right)}{\Gamma \left( -i \mathfrak{a} +ir \mathfrak{b} \right) \Gamma \left( i \mathfrak{a} +ir \mathfrak{b} \right)} \ 2^{-s} \ e^{-\frac{\pi i s}{2}} 
\\[2.5pt]
& \quad \times \int_C dt \ \frac{ \Gamma \left( \frac{1}{2} +i \mathfrak{a} +t \right)  \Gamma \left( \frac{1}{2} -i \mathfrak{a} +t \right) \Gamma \left( -\frac{1}{2} -ir \mathfrak{b} -t \right)}{\Gamma \left( i \mathfrak{a} -ir \mathfrak{b} \right) \Gamma \left( -i \mathfrak{a} -ir \mathfrak{b} \right)} \ 2^{-t} \ e^{\frac{\pi i t}{2}} \ \frac{\Lambda^{3-s-t}}{3-s-t}~,
\end{split}
\end{equation}
where the contour $C$ in the $t$ integration separates the poles of $ \Gamma \left( \frac{1}{2} +i \mathfrak{a} +t \right)  \Gamma \left( \frac{1}{2} -i \mathfrak{a} +t \right)$ from those of $ \Gamma \left( -\frac{1}{2} -ir \mathfrak{b} -t \right)$, and the contour $C'$ for the $s$ integration separates the poles of $ \Gamma \left( \frac{1}{2} -i \mathfrak{a} +s \right)  \Gamma \left( \frac{1}{2} +i \mathfrak{a} +s \right)$ from those of $ \Gamma \left( -\frac{1}{2} +ir \mathfrak{b} -s \right)$. Since the integrand goes to zero as $\Re[t] \rightarrow \infty$, we can take the clockwise contour $C$. Similarly for the contour $C'$. While there exist infinitely many poles at $t = n - 1/2 - i r \mathfrak{b}$ (with $n \geq 0)$ enclosed by $C$ and at $s = n - 1/2 - i r \mathfrak{b}$ (with $n \geq 0)$ enclosed by $C'$, only few poles contribute as $\Lambda \rightarrow \infty$ due to $\Lambda^{3-s-t}$. The poles in the contour $C$ that contribute are at
\begin{equation}
\begin{split}
t = &-\frac{1}{2}-ir \mathfrak{b}, \,  \frac{1}{2}-ir \mathfrak{b}, \,  \frac{3}{2}-ir \mathfrak{b}, \, \frac{5}{2} -ir \mathfrak{b}, \, \frac{7}{2} -ir \mathfrak{b}, \\
& \ 3-s \ \ \left( \text{as} \  3 > \Re [s] \right)~.
\end{split}
\end{equation}
We consider each of them individually and sum them up later.
\begin{enumerate}
\item $ t= -\frac{1}{2}-ir \mathfrak{b}$:\\
The following poles in the contour $C'$,
\begin{equation}
\begin{split}
s = &-\frac{1}{2}+ir \mathfrak{b}, \  \frac{1}{2}+ir \mathfrak{b}, \  \frac{3}{2}+ir \mathfrak{b}, \ \frac{5}{2} +ir \mathfrak{b},\ \frac{7}{2} +ir \mathfrak{b}~,
\end{split}
\end{equation}
 contribute, and the contribution is estimated to be
\begin{equation}
\begin{split}
\sum_r & \Big[ \ \frac{\Lambda^4}{2} +\frac{i}{3} \mu^2 \Lambda^3 -\frac{1}{8} \mu^2 (\mu^2 +2ir \mathfrak{b} +1) \Lambda^2 
\\[2.5pt]
&-\frac{i}{24} \mu^2 (\mu^2+2ir \mathfrak{b}+1)(\mu^2 +4ir \mathfrak{b}+4)\Lambda 
\\[2.5pt]
& -\frac{1}{192} \mu^2  (\mu^2+2ir \mathfrak{b}+1) (\mu^2+4ir \mathfrak{b}+4) (\mu^2+6ir \mathfrak{b}+9)
\\[2.5pt]
& \cdot \left(H_{3-i(\mathfrak{a}-r \mathfrak{b})}+H_{3+i(\mathfrak{a}+r \mathfrak{b})}-\frac{25}{12} -\frac{\pi i}{2} - \ln 2\Lambda - \gamma_E \right)\ \Big]~.
\end{split}
\end{equation}
\item $t= \frac{1}{2}-ir \mathfrak{b}$:\\
The poles inside the contour $C'$,
\begin{equation}
\begin{split}
s = &-\frac{1}{2}+ir\mathfrak{b}, \  \frac{1}{2}+ir\mathfrak{b}, \  \frac{3}{2}+ir\mathfrak{b}, \frac{5}{2} +ir\mathfrak{b}~,
\end{split}
\end{equation}
leads to the contribution,
\begin{equation}
\begin{split}
\sum_r & \  \mu^2  \Big[ \ -\frac{i}{3} \Lambda^3 +\frac{1}{4} \mu^2 \Lambda^2 +\frac{i}{8} \mu^2 (\mu^2+2ir\mathfrak{b}+1)\Lambda 
\\[2.5pt]
& +\frac{1}{48} \mu^2  (\mu^2+2ir\mathfrak{b}+1) (\mu^2+4ir\mathfrak{b}+4)
\\[2.5pt] 
&\times \left(H_{2-i(\mathfrak{a}-r\mathfrak{b})}+H_{2+i(\mathfrak{a}+r\mathfrak{b})}-\frac{11}{6} -\frac{\pi i}{2} - \ln 2\Lambda - \gamma_E \right) \ \Big]~.
\end{split}
\end{equation}
\item $ t= \frac{3}{2}-ir\mathfrak{b}$:\\
The following poles,
\begin{equation}
\begin{split}
s= &-\frac{1}{2}+ir\mathfrak{b}, \  \frac{1}{2}+ir\mathfrak{b}, \  \frac{3}{2}+ir\mathfrak{b} ~
\end{split}
\end{equation}
gives the contribution,
\begin{equation}
\begin{split}
-\frac{1}{2}  & \sum_r \ \mu^2 (\mu^2-2ir\mathfrak{b}+1) \Big[ \ \frac{\Lambda^2}{4} +\frac{i}{4} \mu^2 \Lambda 
\\[2.5pt]
& +\frac{1}{16} \mu^2  (\mu^2+2ir\mathfrak{b}+1)\left(H_{1-i(\mathfrak{a}-r\mathfrak{b})}+H_{1+i(\mathfrak{a}+r\mathfrak{b})}-\frac{3}{2} -\frac{\pi i}{2} - \ln 2\Lambda - \gamma_E \right) \Big]~,
\end{split}
\end{equation}
\item $t= \frac{5}{2}-ir\mathfrak{b}$:\\
The contribution from poles,
\begin{equation}
\begin{split}
s =& -\frac{1}{2}+ir\mathfrak{b}, \  \frac{1}{2}+ir\mathfrak{b}
\end{split}
\end{equation}
is estimated to be
\begin{equation}
\begin{split}
\frac{1}{6} & \sum_r \ \mu^2 (\mu^2-2ir\mathfrak{b}+1)(\mu^2-4ir\mathfrak{b}+4) 
\\[2.5pt] 
& \times \left[ \ \frac{i}{4}\Lambda+\frac{1}{8} \mu^2 \left(H_{-i(\mathfrak{a}-r\mathfrak{b})}+H_{i(\mathfrak{a}+r\mathfrak{b})}-1 -\frac{\pi i}{2} - \ln 2\Lambda - \gamma_E \right) \right]~,
\end{split}
\end{equation}
\item $t= \frac{7}{2}-ir\mathfrak{b}$:\\
In this case, only one pole,
\begin{equation}
\begin{split}
s = & -\frac{1}{2}+ir\mathfrak{b}
\end{split}
\end{equation}
gives the contribution which is given by
\begin{equation}
\begin{split}
-\frac{1}{192} & \sum_r \ \mu^2 (\mu^2-2ir\mathfrak{b}+1)(\mu^2-4ir\mathfrak{b}+4)(\mu^2-6ir\mathfrak{b}+9) 
\\[2.5pt] 
& \times \left( \ H_{-1-i(\mathfrak{a}-r\mathfrak{b})}+H_{-1+i(\mathfrak{a}+r\mathfrak{b})}-\frac{\pi i}{2} - \ln 2\Lambda - \gamma_E \right)~.
\end{split}
\end{equation}
\end{enumerate}
Summing up over the contributions from the above five poles at $ t= -1/2-ir\mathfrak{b}, \,  1/2 -ir\mathfrak{b}, \,  3/2-ir\mathfrak{b}, \, 5/2 -ir\mathfrak{b}, \, 7/2 -ir\mathfrak{b} $, we obtain 
\begin{equation}\label{app:eq:3rd:5poles}
\begin{split}
& \Lambda^4 -\frac{1}{2} \mu^2 \Lambda^2 +\frac{3}{4}\mu^2(\mu^2-4\mathfrak{b}^2+1)(\ln 2\Lambda +\gamma) \\
& -\frac{3}{8} \mu^2 (\mu^2-4\mathfrak{b}^2+1) \left(H_{-i(\mathfrak{a}-\mathfrak{b})}+H_{-i(\mathfrak{a}+\mathfrak{b})}+H_{i(\mathfrak{a}-\mathfrak{b})}+H_{i(\mathfrak{a}+\mathfrak{b})} \right) \\
& -\mathfrak{b}^4+\frac{1}{96}\mathfrak{b}^2(35\mu^4+228\mu^2+264)-\frac{1}{384}\mu^2(\mu^6+38\mu^4-23\mu^2+228) ~.
\end{split}
\end{equation}
Finally we check the contribution from the pole at $ t= 3-s$. It has non-zero value at $3>\Re [s] $, otherwise the pole $t=3-s$ cannot be enclosed by the contour $C$. Performing the integration of Eq.~(\ref{app:eq:rho:3rd}) over $t$ and using $\Gamma(z)\Gamma(1-z)=\pi / \sin(\pi z)$, we obtain 
\begin{equation}\label{app:eq:gh:step1}
\begin{split}
& -\frac{\mu^2}{16}\sum_r e^{-\pi r \mathfrak{b}} \int_{C'} ds \ \frac{e^{- \pi i s} \cdot \sinh \left[ \pi (\mathfrak{a} -r\mathfrak{b}) \right]  \sinh \left[ \pi (\mathfrak{a}+r\mathfrak{b}) \right] }{\sin \left[ \pi \left( \frac{1}{2}-i\mathfrak{a} +s \right) \right] \sin \left[ \pi \left( \frac{1}{2}+i\mathfrak{a} +s \right) \right] \sin \left[ \pi \left( \frac{1}{2} -ir\mathfrak{b} +s \right) \right] }  
\\[3pt]
& \times  \ \frac{ \left( \frac{5}{2} +i\mathfrak{a} -s \right)\left( \frac{5}{2} -i\mathfrak{a} -s \right)\left( \frac{3}{2} +i\mathfrak{a} -s \right)\left( \frac{3}{2} -i\mathfrak{a} -s \right)\left( \frac{1}{2} +i\mathfrak{a} -s \right)\left( \frac{1}{2} -i\mathfrak{a} -s \right) }{\left( -\frac{7}{2} -ir\mathfrak{b} +s \right)\left( -\frac{5}{2} -ir\mathfrak{b} +s \right)\left( -\frac{3}{2} -ir\mathfrak{b} +s \right)\left( -\frac{1}{2} -ir\mathfrak{b} +s \right)\left( -\frac{1}{2} +ir\mathfrak{b} -s \right)} ~.
\end{split}
\end{equation}
The evaluation of the above integration can be done using a similar trick used in the calculation of the backreaction~\cite{Adshead:2018oaa}. The expression in Eq.~(\ref{app:eq:gh:step1}) can be rewritten as
\begin{equation}\label{app:eq:gh:step2}
\begin{split}
& -\frac{\mu^2}{16}\sum_r e^{-\pi r \mathfrak{b}} \int_{C'} ds \ \frac{e^{- \pi i s} \cdot \sinh \left[ \pi (\mathfrak{a} -r\mathfrak{b}) \right]  \sinh \left[ \pi (\mathfrak{a}+r\mathfrak{b}) \right] }{\sin \left[ \pi \left( \frac{1}{2}-i\mathfrak{a} +s \right) \right] \sin \left[ \pi \left( \frac{1}{2}+i\mathfrak{a} +s \right) \right] \sin \left[ \pi \left( \frac{1}{2} -ir\mathfrak{b} +s \right) \right] }  \\
& \quad \quad \quad \times  \Big[ g(s) -g(s-1) +h(s) \Big]~,
\end{split}
\end{equation}
where the functions $g(s)$ and $h(s)$ are given by,
\begin{equation}
\begin{split}
& g(s) = c_1 \left( -\frac{1}{2}-ir\mathfrak{b} +s \right) +c_2 \left( -\frac{1}{2}-ir\mathfrak{b} +s \right)^2 + \frac{ c_{-1}}{ -\frac{1}{2}-ir\mathfrak{b} +s } \\
& \quad + \frac{ {c'}_{-1}}{ -\frac{1}{2}+ir\mathfrak{b} -s }+ \frac{ c_{-3}}{ -\frac{3}{2}-ir\mathfrak{b} +s } + \frac{ c_{-5}}{ -\frac{5}{2}-ir\mathfrak{b} +s } \\
& h(s) = \frac{ c_{-7}}{ -\frac{7}{2}-ir\mathfrak{b} +s },
\end{split}
\end{equation}
with the coefficients given below:
\begin{equation}
\begin{split}
& c_1 = \frac{1}{2}-6ir\mathfrak{b}~, \\[3pt]
& c_2 = -\frac{1}{2}~, \\[3pt]
& c_{-1} = \frac{1}{8} [-12-11\mathfrak{a}^2+2\mathfrak{a}^4+\mathfrak{a}^6 +44ir\mathfrak{b}+16ir\mathfrak{a}^2\mathfrak{b}-4ir\mathfrak{a}^4\mathfrak{b}+59\mathfrak{b}^2-3\mathfrak{a}^4\mathfrak{b}^2 -32ir\mathfrak{b}^3 \\[3pt]
& \quad \quad \quad \quad +8ir\mathfrak{a}^2\mathfrak{b}^3 -2\mathfrak{b}^4 +3\mathfrak{a}^2\mathfrak{b}^4-4ir\mathfrak{b}^5 -\mathfrak{b}^6 ]~, \\[3pt]
& {c'}_{-1} = \frac{1}{24} [36 +49\mathfrak{a}^2 +14\mathfrak{a}^4+\mathfrak{a}^6 -132ir\mathfrak{b} -96ir\mathfrak{a}^2\mathfrak{b} -12ir\mathfrak{a}^4\mathfrak{b} -193\mathfrak{b}^2 -72\mathfrak{a}^2\mathfrak{b}^2 \\[3pt]
& \quad  \quad  \quad  \quad -3\mathfrak{a}^4\mathfrak{b}^2 +144ir\mathfrak{b}^3 +24ir\mathfrak{a}^2\mathfrak{b}^3+58\mathfrak{b}^4+3\mathfrak{a}^2\mathfrak{b}^4 -12ir\mathfrak{b}^5 -\mathfrak{b}^6]~, \\[3pt]
& c_{-3} = \frac{1}{8} [-12-13\mathfrak{a}^2 -2\mathfrak{a}^4 -\mathfrak{a}^6 +44ir\mathfrak{b} +16ir\mathfrak{a}^2\mathfrak{b} -4ir\mathfrak{a}^4\mathfrak{b} +61\mathfrak{b}^2 +3\mathfrak{a}^4\mathfrak{b}^2-32ir\mathfrak{b}^3 \\[3pt]
& \quad \quad  \quad  \quad  +8ir\mathfrak{a}^2\mathfrak{b}^3+2\mathfrak{b}^4  -3\mathfrak{a}^2\mathfrak{b}^4-4ir\mathfrak{b}^5+\mathfrak{b}^6 ]~, \\[3pt]
& c_{-5} = \frac{1}{24} [ -36 -23 \mathfrak{a}^2 +14\mathfrak{a}^4 +\mathfrak{a}^6 +132ir\mathfrak{b} +96ir\mathfrak{a}^2\mathfrak{b} +12ir\mathfrak{a}^4\mathfrak{b} +167\mathfrak{b}^2 -72\mathfrak{a}^2\mathfrak{b}^2 \\[3pt]
& \quad \quad  \quad  \quad  -3\mathfrak{a}^4\mathfrak{b}^2-144ir\mathfrak{b}^3 -24ir\mathfrak{a}^2\mathfrak{b}^3+58\mathfrak{b}^4+3\mathfrak{a}^2\mathfrak{b}^4+12ir\mathfrak{b}^5-\mathfrak{b}^6]~, \\[3pt]
& c_{-7} = -3-3\mathfrak{a}^2+15\mathfrak{b}^2 ~.
\end{split}
\end{equation}
The first two terms in Eq.~(\ref{app:eq:gh:step2}) can be combined into one contour integration such that a new contour $C''$ encloses only three poles at $ s= -1/2 +i\mathfrak{a},\ - 1/2-i\mathfrak{a},\ -3/2 +ir\mathfrak{b}$:
\begin{equation}\label{app:eq:gh:step3}
\begin{split}
 -\frac{\mu^2}{16}\sum_r & e^{-\pi r \mathfrak{b}} \left [ \int_{C''} ds \ \frac{e^{- \pi i s} \cdot \sinh \left[ \pi (\mathfrak{a} -r\mathfrak{b}) \right]  \sinh \left[ \pi (\mathfrak{a}+r\mathfrak{b}) \right] \cdot g(s) }{\sin \left[ \pi \left( \frac{1}{2}-i\mathfrak{a} +s \right) \right] \sin \left[ \pi \left( \frac{1}{2}+i\mathfrak{a} +s \right) \right] \sin \left[ \pi \left( \frac{1}{2} -ir\mathfrak{b} +s \right) \right] } \right .  \\[4pt]
& \left .  +\int_{C'} ds \ \frac{e^{- \pi i s} \cdot \sinh \left[ \pi (\mathfrak{a} -r\mathfrak{b}) \right]  \sinh \left[ \pi (\mathfrak{a}+r\mathfrak{b}) \right] \cdot h(s) }{\sin \left[ \pi \left( \frac{1}{2}-i\mathfrak{a} +s \right) \right] \sin \left[ \pi \left( \frac{1}{2}+i\mathfrak{a} +s \right) \right] \sin \left[ \pi \left( \frac{1}{2} -ir\mathfrak{b} +s \right) \right] } \ \right ]~.
\end{split}
\end{equation}
The first integration in Eq.~(\ref{app:eq:gh:step3}) is straightforward to evaluate, and it is given by
\begin{equation}\label{app:eq:1st:piece}
\begin{split}
 \frac{\mu^2}{8} \sum_r & \left [ -e^{\pi (\mathfrak{a}-r\mathfrak{b})} \frac{\sinh \left[ \pi (\mathfrak{a}+r\mathfrak{b}) \right] }{\sinh(2 \pi \mathfrak{a}) } \ g \left( -\frac{1}{2}+i\mathfrak{a} \right) \right .  
 \\[3pt]
&  \left . -e^{-\pi (\mathfrak{a}+r\mathfrak{b})} \frac{\sinh \left[ \pi (\mathfrak{a}-r\mathfrak{b}) \right] }{\sinh(2 \pi \mathfrak{a}) } \ g \left( -\frac{1}{2}-i\mathfrak{a} \right) +g \left(-\frac{3}{2} +ir\mathfrak{b} \right) \ \right ]~.
\end{split}
\end{equation}
For the second integration in Eq.~(\ref{app:eq:gh:step3}), we take a counter-clockwise contour $C'$ as the integrand with $h(s)$ wildly oscillates as $ \Re [s] \rightarrow -\infty$. An infinite number of poles of $s$ are enclosed by the contour $C'$, namely poles at $s = n -1/2+i\mathfrak{a}$, $n -1/2 - i\mathfrak{a}$, and $n -3/2 + ir\mathfrak{b}$ (with $n\leq 0$). The second integration in Eq.~(\ref{app:eq:gh:step3}) is estimated to be
\begin{equation}\label{app:eq:2nd:piece}
\begin{split}
& \frac{\mu^2}{8} \sum_r 3(1+\mathfrak{a}^2-5\mathfrak{b}^2) \Big[ e^{\pi (\mathfrak{a}-r\mathfrak{b})} \ \frac{\sinh \left[ \pi (\mathfrak{a}+r\mathfrak{b}) \right]}{\sinh (2 \pi \mathfrak{a} )} H_{3-i(\mathfrak{a}-r\mathfrak{b})} \\[3pt]
& \quad \quad \quad \quad \quad \quad \quad \quad \quad \quad \quad+e^{-\pi (\mathfrak{a}+r\mathfrak{b})} \ \frac{\sinh \left[ \pi (\mathfrak{a}-r\mathfrak{b}) \right]}{\sinh (2 \pi \mathfrak{a} )}H_{3+i(\mathfrak{a}+r\mathfrak{b})} -\frac{25}{12} \Big]~.
\end{split}
\end{equation}
Summing over two contributions in Eq.~(\ref{app:eq:1st:piece}) and~(\ref{app:eq:2nd:piece}), we obtain the contribution from the pole at $t= 3-s $:
\begin{equation}\label{app:eq:3rd:lastpole}
\begin{split}
& \frac{1}{384} \Big[192 \mathfrak{b}^4 +\mu^2 (1+\mu^2) (-228+37\mu^2+\mu^4 )-4\mathfrak{b}^2 (132-156 \mu^2 +35\mu^4 ) \\
& \quad  \quad +48\mathfrak{a}\mathfrak{b} (-26\mu^2+4\mathfrak{b}^2-11) \sinh(2 \pi \mathfrak{b}) \text{csch}( 2 \pi \mathfrak{a} ) \ \Big] \\
& +\frac{3}{16} \mu^2 (\mu^2 -4\mathfrak{b}^2 +1 ) \Big[ \left(1+ \sinh(2 \pi \mathfrak{b}) \text{csch}( 2 \pi \mathfrak{a} ) \right) \left( H_{i(\mathfrak{a}-\mathfrak{b})} +H_{-i(\mathfrak{a}-\mathfrak{b})} \right) \\
& \quad \quad \quad \quad \quad \quad \quad \quad \quad+ \left(1- \sinh(2 \pi \mathfrak{b}) \text{csch}( 2 \pi \mathfrak{a} ) \right) \left( H_{i(\mathfrak{a}+\mathfrak{b})} +H_{-i(\mathfrak{a}+\mathfrak{b})} \right) \Big]~.
\end{split}
\end{equation}
Therefore, the third term of the second line in Eq.~(\ref{app:eq:rhot:exp2}) is obtained by summing over Eqs~(\ref{app:eq:3rd:5poles}) and~(\ref{app:eq:3rd:lastpole}):
\begin{equation}
\begin{split}
\sum_r \int^\Lambda_0 dx \, x^2 | s_r |^2  =&
 \ \Lambda^4 -\frac{1}{2} \mu^2 \Lambda^2 +\frac{3}{4} \mu^2 ( \mu^2-4\mathfrak{b}^2 +1)( \ln 2\Lambda + \gamma_E ) \\[3pt]
& +\frac{1}{16} \left[-8\mathfrak{b}^4 +22 \mathfrak{b}^2 +64 \mu^2 \mathfrak{b}^2-19\mu^2 -7\mu^4  \right] \\[3pt]
& +\frac{1}{8} \mathfrak{a}\mathfrak{b} (-26\mu^2 +4\mathfrak{b}^2 -11) \ \sinh (2 \pi \mathfrak{b}) \text{csch} (2 \pi \mathfrak{a} ) \\[3pt]
& -\frac{3}{16} \mu^2 (\mu^2 -4\mathfrak{b}^2 +1) \Big[ \left(1- \sinh(2 \pi \mathfrak{b}) \text{csch}( 2 \pi \mathfrak{a} ) \right) \left( H_{i(\mathfrak{a}-\mathfrak{b})} +H_{-i(\mathfrak{a}-\mathfrak{b})} \right) \\[3pt]
& \quad \quad \quad + \left(1+ \sinh(2 \pi \mathfrak{b}) \text{csch}(2 \pi \mathfrak{a} ) \right) \left( H_{i(\mathfrak{a}+\mathfrak{b})} +H_{-i(\mathfrak{a}+\mathfrak{b})} \right) \Big]~.
\end{split}
\end{equation}
So far we have completed the calculation of three terms in the second line in Eq.~(\ref{app:eq:rhot:exp2}), and therefore, the fermion energy density as a function of the conformal time is given by
\begin{equation}
\begin{split}
\rho_\psi(\tau) =  \ \frac{4 \pi}{\tau^4} & \Big[ \ \frac{1}{4} \mu^4 \ln \frac{\mu}{2\Lambda} +\frac{1}{4}\mu^2 (\mu^2 +4\mathfrak{b}^2 +1) \left( \ln 2\Lambda +\gamma_E \right) \\ 
& -\frac{13}{16} \mu^2 -\mu^2 \mathfrak{b}^2 -\frac{11}{8} \mathfrak{b}^2 +\frac{1}{2}\mathfrak{b}^4 +\frac{1}{8} \mathfrak{a}\mathfrak{b} (2\mu^2-4\mathfrak{b}^2+11) \sinh(2\pi \mathfrak{b}) \text{csch} (2 \pi \mathfrak{a}) 
\\[3.5pt]
& -\frac{1}{16} \mu^2 (\mu^2+4\mathfrak{b}^2 +12i\mathfrak{b} +1) \left( 1- \sinh(2\pi \mathfrak{b})\text{csch}  (2 \pi \mathfrak{a})  \right) H_{-i(\mathfrak{a}-\mathfrak{b})} 
\\[3.5pt]
& -\frac{1}{16} \mu^2 (\mu^2+4\mathfrak{b}^2 -12i\mathfrak{b} +1) \left( 1- \sinh(2\pi \mathfrak{b})\text{csch} (2 \pi \mathfrak{a}) \right) H_{i(\mathfrak{a}-\mathfrak{b})} 
\\[3.5pt]
& -\frac{1}{16} \mu^2 (\mu^2+4\mathfrak{b}^2 +12i\mathfrak{b} +1) \left( 1+  \sinh(2\pi \mathfrak{b})\text{csch}  (2 \pi \mathfrak{a}) \right) H_{i(\mathfrak{a}+\mathfrak{b})} 
\\[3.5pt]
& -\frac{1}{16} \mu^2 (\mu^2+4\mathfrak{b}^2 -12i\mathfrak{b} +1) \left( 1+ \sinh(2\pi \mathfrak{b})\text{csch}  (2 \pi \mathfrak{a}) \right) H_{-i(\mathfrak{a}+\mathfrak{b})} \ \Big]~,
\end{split}
\end{equation}
where $\mathfrak{a} = \sqrt{\mu^2 + 4\xi^2}$ and $\mathfrak{b}=2\xi$. The fermion energy density as a function of the cosmic time $t$ is obtained as 
\begin{equation}
  \rho_\psi (t) = \frac{\rho_\psi (\tau)}{a^4}~.
\end{equation}
One notices that the quartic and quadratic divergences were cancelled in $\rho_\psi$, and the final result has only logarithmic divergence.
Similarly to the case of the backreaction, we simply drop the logarithmic divergence and consider the finite terms. When $\mu^2 \ll \xi$ and $ 1 \ll \xi $, the fermion energy density is approximated by
\begin{equation}
\rho_\psi(t) \approx \frac{16\pi^2}{a^4\tau^4} \mu^2 \xi^3~.
\end{equation}
Just like the case of the backreaction, the condition $\mu \ll 1$ is not necessary to derive above approximation.

%%%%%%%%%%%%%%%%%%%%%%%%%%%%%%%%%%%%%%%%
%%%%%%%%%%%%%%%%%%%%%%%%%%%%%%%%%%%%%%%%
%%%%%%%%%%%%%%%%%%%%%%%%%%%%%%%%%%%%%%%%
%%%%%%%%%%%%%%%%%%%%%%%%%%%%%%%%%%%%%%%%
%%%%%%%%%%%%%%%%%%%%%%%%%%%%%%%%%%%%%%%%
%%%%%%%%%%%%%%%%%%%%%%%%%%%%%%%%%%%%%%%%

%%%%%%%%%%%%%%%%%
% References
%%%%%%%%%%%%%%%%%
\bibliography{lit}

\end{document}